\documentclass[12pt]{article} 

\usepackage{amsmath}
\usepackage{amsfonts}
\usepackage{times}
\usepackage{epsf}

\def\({\left(}
\def\){\right)}
\def\[{\left[}
\def\]{\right]}
\def\non{ \nonumber }
\def\b{\beta}
\def\d{\delta}
\def\a{\alpha}
\def\l{\lambda}
\def\la{\lambda}
\def\e{\epsilon}
\def\s{\sigma}
\def\half{\textstyle{\frac 1 2}}
\begin{document} 
\phantom{a}

$$ $$
\vskip 0.5cm

\centerline{\bf \large 
New formulae for solutions of quantum Knizhnik-Zamolodchikov}
\centerline{\bf \large
equations on level -4 and correlation functions}
\vskip 2cm
\centerline{Hermann Boos
\footnote{on leave of absence from the Institute for High Energy Physics,
Protvino, 142284, Russia}}
\centerline{\it Max-Planck Institut f{\"u}r Mathematik}
\centerline{\it Vivatsgasse 7, 53111 Bonn, Germany} 
\vskip 0.5cm
\centerline{Vladimir Korepin}
\centerline{\it C.N.~Yang Institute for Theoretical Physics}
\centerline{\it State University of New York at Stony Brook}
\centerline{\it Stony Brook, NY 11794--3840, USA}
\vskip 0.5cm
\centerline{Feodor  Smirnov
\footnote{Membre du CNRS}}
\centerline{\it LPTHE, Tour 16, 1-er {\'e}tage, 4, pl. Jussieu}
\centerline{\it 75252, Paris Cedex 05, France}
$$ $$
\vskip 1.5cm
\begin{abstract}
This paper is continuation of our previous papers \cite{bks} and \cite{bks1}.
We discuss in more detail 
a new form of solution to the quantum Knizhnik-Zamolodchikov 
equation [qKZ] on level $-4$ obtained in the paper \cite{bks1}
for the Heisenberg XXX spin chain. The main advantage of this
form is it's explicit reducibility to one-dimensional integrals.
We argue that the deep mathematical reason for this is some
special cohomologies of deformed Jacobi varieties. 
We apply this new form of solution
to the correlation functions using the Jimbo-Miwa conjecture \cite{JM}.
A formula (\ref{cor2}) for the correlation functions obtained in this way
is in a good agreement with the ansatz for the emptiness formation
probability from the paper \cite{bks}.
Our previous conjecture on a structure of correlation functions of
the XXX model in the homogeneous limit through the Riemann zeta functions
at odd arguments is a corollary of the formula (\ref{cor2}).
\end{abstract}

\newpage
\noindent
\section{ Introduction.}

In the paper \cite{bks1} we have suggested
a new form of solution to the quantum Knizhnik-Zamolodchikov 
equation [qKZ] on level -4 \cite{FR,JM}. The main idea was to
use a duality between the solutions to the qKZ on level -4 and level 0
which first appeared in a context of the form-factors of integrable 
models of quantum field theory \cite{book}. 
The latter solutions 
have much more simple structure in comparison with the qKZ solution on level -4
because they all can be reduced to single integrals. 
The main point is to get solutions to the qKZ on level -4 from the 
known qKZ solutions on level 0 inverting some special matrix
$\mathcal F$ of dimension $\binom{2n}{n}-\binom{2n}{n-1}$  
built up from different level 0 solutions \cite{book,count}. 
The matrix may be represented
as a product of two matrices $\mathcal P$ and $\mathcal H$ which carry
respectively transcendental and rational dependence on the rapidities
$\b_1,\ldots,\b_{2n}$. 
It is possible to invert the transcendental part   
using some facts about deformed hyper-elliptic integrals, in particular,
the deformed Riemann bilinear relation \cite{book,SKyo,sm1986}.
 But the problem to invert the rational
part appeared to be rather complicated. In order to do this we need 
a theorem about a special form of solutions to the qKZ on level -4, 
namely, an ansatz given by the formula (20) from \cite{bks1}.
We have proved that the functions $\widetilde h$ that appear in
the above formula are polynomials of all arguments. We know how 
these polynomials look like for few first cases. Unfortunately,
an explicit form of these polynomials in general case is still to
be found. 
In this paper we are trying to make some steps in this direction,
namely, we have found a representation for the polynomials $\widetilde h$ 
in terms of contour integrals of a special form. The formula obtained
has a cohomological meaning. In some sense it corresponds to 
a deformation of the cohomologies described by the theorem 
from the sixth section of \cite{bks1} conjectured in  \cite{ns} and
proved by Nakayashiki in \cite{n}. 

The next important issue is to apply the solution to the qKZ
on level -4 to the correlation functions of the Heisenberg XXX spin chain.
This is possible due to the nice conjecture suggested by Jimbo and Miwa 
in 1996 \cite{JM}. This conjecture says that there is a direct 
connection of the solution to the qKZ on level -4 with
any correlation function for the Heisenberg XXZ model in the massless
regime and, in particular, for the XXX model.
The integral representation that follows from this conjecture was later 
confirmed by the algebraic Bethe ansatz approach \cite{Maillet1},
\cite{Maillet2}.
Our main statement is that using the Jimbo-Miwa conjecture together with our
formulae for the solution to the qKZ on level -4 we come to the result
that any correlation function of the XXX model can be expressed through
the function $G$ defined by the formulae (3.6) and
(3.13) of the paper \cite{bks} with coefficients being rational 
functions like in the ansatz for the emptiness formation probability
given by the formula (3.20) of the above mentioned paper.
Our previous conjecture \cite{bk1,bk2,BKNS} 
that any correlation function in the 
homogeneous limit can be expressed in terms of the
Riemann $\zeta$-function at odd arguments with rational coefficients
follows from the above statement which is valid for more general case 
when the spectral parameters are different.

The paper is organized as follows. In Section 2 we discuss in more detail
the new form of solutions to the qKZ on level -4 which is similar to the
Smirnov's solution to the qKZ on level 0, namely, it also has a property
of reducibility to one-dimensional integrals. In Section 3 we apply 
this solution to the correlation functions of the XXX model using the
Jimbo-Miwa conjecture. Some particular cases are discussed in detail
in the Appendix.

\section{Solutions to qKZ on level -4}
\vskip 0.5cm
Let us remind the reader the main statement of our previous
work \cite{bks1}, namely, the formula from the theorem of the
section 5 for the solutions to the qKZ on level -4
belonging to singlet subspace of $(\mathbb{C}^2)^{\otimes 2n}$
are
counted by integers $\{k_1,\cdots ,k_{n-1}\}$, with $|k_j|\le n-1$, $\forall j$
\begin{align}
&g_{\{k_1,\cdots ,k_{n-1}\}}(\b _1,\cdots, \b _{2n})=
e^{\half\sum \b _j}
\prod\limits _{i<j}\frac 1 {\zeta (\b _i-\b _j)}
\int\limits _{-\infty}^{\infty} d\a _1\cdots \int\limits _{-\infty}^{\infty} 
d\a _{n-1}\prod\limits _{i,j}\varphi (\a _i -\b _j)\non\\
&\times
\text{det}|e^{k_i\a _j}|_{1\le i,j\le n-1}\
\ \widetilde{h}(\a _1,\cdots \a _{n-1}|\b _1,\cdots ,\b _{2n})
\label{theor}
\end{align}
where
\begin{align}
\varphi (\a)=\Gamma \(\frac 1 4 +\frac {\a}{2\pi i}\)
\Gamma \(\frac 1 4 -\frac {\a}{2\pi i}\)
\label{phi}
\end{align}
\begin{align}
\zeta (\b)=
\exp\(-\int\limits _{0}^{\infty}
\frac {\sin ^2\frac 1 2(\b+\pi i)k\ e^{-\frac {\pi k}2}}
{k\sinh (\pi k)\cosh \(\frac {\pi k}2\)}\)
\label{zeta}
\end{align}
and $\widetilde{h}(\a _1,\cdots \a _{n-1}|\b _1,\cdots ,\b _{2n})$
is a polynomial of all its arguments taking values in
$(\mathbb{C}^2)^{\otimes 2n}$ which is skew-symmetric with
respect to $\a _1,\cdots \a _{n-1}$.


As was explained in \cite{bks1} one can present $\widetilde{h}$ in the
following form:
\begin{align}
&\widetilde{h}(\a _1, \cdots ,\a _{n-1}|\b _1,\cdots ,\b _{2n})=
\sum\limits
_{\{1,\cdots,2n\}=\{i_1,\cdots ,i_n\}\cup \{j_{1},\cdots ,j_{n}\}}
v(\a _1,\cdots \a _{n-1}|
\b _{i_1},\cdots ,\b _{i_n}|\b _{j_1},\cdots ,\b _{j_n})
\non\\&\times
\prod\limits _{p,q=1}^n \frac {\b _{i_p}-\b _{j_q}+\pi i} 
{\b _{i_p}-\b _{j_q}}
\ w^{\dag}_{\epsilon _1,\cdots,\epsilon _{2n}}(\b _1,\cdots\b_{2n})&\non
\end{align}
where $\e_{i_p}=-$ and $\e_{j_q}=+$. The
vectors $w^{\dag}_{\epsilon _1,\cdots,\epsilon _{2n}}(\b _1,\cdots\b_{2n})$ 
constitute a special basis in the weight zero subspace of 
$(\mathbb{C}^2)^{\otimes 2n}$. Taking components of these
vectors with respect to the natural basis of the tensor product
one obtains the matrix
$$w^{\dag}\ _{\epsilon _1,\cdots,\epsilon _{2n}}
^{\epsilon _1',\cdots,\epsilon _{2n}'}
(\b _1,\cdots\b_{2n})
$$
which is triangular with respect to certain ordering.

In order to find the function $v$ we should solve two sets of equations.
The first one follows
from requirement that $\widetilde{h}$ must belong to singlet subspace:
\begin{align}
\sum\limits _{p=1}^{n+1}v(\a _1,\cdots ,\a _{n-1}|
\b _{i_1},\cdots ,\b _{i_{n-1}},\b _{j_p}|\b_{j_1},
\cdots ,\widehat {\b  _{j_p}},\cdots ,\b _{j_{n+1}})
\prod\limits
_{q\ne p}\frac {\b _{j_p}-\b _{j_q} -\pi i}{\b _{j_p}-\b _{j_q}}=0
\label{singlet}
\end{align}
The second equation is equivalent to the fact that $\widetilde{h}$
is obtained by inverting the matrix $\mathcal{H}$:
\begin{align}
&\sum\limits
_{\{1,\cdots,2n\}=\{i_1,\cdots ,i_n\}\cup \{j_{1},\cdots ,j_{n}\}}
v(\a _1,\cdots ,\a _{n-1}|
\b _{j_1},\cdots ,\b _{j_n}|\b _{i_1},\cdots ,\b _{i_n})
\non\\
&
\times u(\a _1',\cdots ,\a _{n-1}'|
\b _{i_1},\cdots ,\b _{i_n}|
\b _{j_1},\cdots ,\b _{j_n})\prod
\limits _{p,q=1}^n\frac 1 {\b _{i_p}-\b _{j_q}}=
c(\a _1,\cdots ,\a _{n-1}|\a _1',\cdots ,\a _{n-1}')
\label{intersect}
\end{align}
where 
\begin{align}
u(\a _1,\cdots \a _{n-1}|\b _1,\cdots ,\b _{n}|\b _{n+1},\cdots ,\b _{2n})=
\text{det}(A_i(\a _j|\b _1,\cdots ,\b _{n}|\b _{n+1},\cdots ,\b _{2n}))
|_{i,j=1,\cdots , n-1}
\end{align}
and the polynomials $A_i(\a)$ depend on $\b _j$ as parameters and may be
defined through the generating function:
\begin{align}
&\sum\limits _{i=1}^{n-1}\gamma ^{n-i-1}
A_i(\a|\b _1,\cdots ,\b _{n}|\b _{n+1},\cdots ,\b _{2n})=
\frac {\prod\limits _{j=1}^{2n}
(\a -\b _j+\frac {\pi i}2)}{\a -\gamma +\pi i}-
\frac {\prod\limits _{j=1}^{2n}(\a -\b _j-\frac {\pi i}2)}
{\a -\gamma -\pi i}+\non\\
&+\frac {\pi i\prod\limits _{j=1}^{n}(\a -\b _j-\frac {\pi i}2)
(\gamma-\b _{n+j}+\frac {\pi i}2)}
{(\a-\gamma)(\a -\gamma -\pi i)}+
\frac {\pi i\prod\limits _{j=1}^{n}(\gamma -\b _j-\frac {\pi i}2)
(\a-\b _{n+j}+\frac {\pi i}2)}
{(\a-\gamma)(\a -\gamma +\pi i)}
\label{gener}
\end{align}
The function
$c(\a _1,\cdots \a _{n-1}|\a _1',\cdots \a _{n-1}')$
is the "intersection form". Essential part of this "intersection form" is 
$\text{det}\left|c(\a _i,\a '_j)\right|$ with 
\begin{align}
c(\a _1,\a _2)=\frac {\prod\limits _{j=1}^{2n}
(\a _1 -\b _j+\frac {\pi i}2)}{\a _1-\a _2+\pi i}-
\frac {\prod\limits _{j=1}^{2n}
(\a _1-\b _j-\frac {\pi i}2)}{\a _1-\a _2-\pi i}-
\frac {\prod\limits _{j=1}^{2n}
(\a _2 -\b _j+\frac {\pi i}2)}{\a _2-\a _1+\pi i}+
\frac {\prod\limits _{j=1}^{2n}
(\a _2-\b _j-\frac {\pi i}2)}{\a _2-\a _1-\pi i}
\label{c}
\end{align}
The rigorous definition of the "intersection form" was given in 
the end of section 5 of \cite{bks1}. Now we do not need it.

Let us consider 
$v(\a _1,\cdots \a _{n-1}|
\b _{i_1},\cdots ,\b _{i_n}|\b _{j_1},\cdots ,\b _{j_n})$
for different partitions as $ \binom{2n}{n}$ independent unknowns. 

Actually, we have $\binom{2n}{n-1}$ equations from (\ref{singlet}). 
In order to analyze the number of equations which follow from
the second requirement (\ref{intersect}) we need the formula (17)
from \cite{bks1}
\begin{align}
A_{k}(\a|\b _{i_1},\cdots ,\b_{i_n}|\b _{j_1},\cdots ,\b_{j_n})=
s_k(\a )+\sum\limits _{l=1}^{n-1}c_{kl}
(\b _{i_1},\cdots ,\b_{i_n}|\b _{j_1},\cdots ,\b_{j_n})s_{-l}(\a)\label{As}
\end{align}
where 
\begin{align}
s_k(\a )=A_k(\a |\b _1,\cdots ,\b _n|\b _{n+1},\cdots ,\b _{2n}),\quad
s_{-k}(\a )=\a ^{n-k-1},\quad k=1,\ldots n-1
\label{s}
\end{align}
and the matrix $c_{kl}$ is symmetric. 
The formula (\ref{As}) follows from the fact that
\begin{align}
c(\a _1,\a _2)=\sum\limits_{k=1}^{n-1}
\(s_k(\a _1)s_{-k}(\a_2)-s_k(\a_2)s_{-k}(\a_1)\)
\label{ss}
\end{align}
and due to (\ref{c}) it is symmetric function of all $2n$ variables
$\b_1,\ldots,\b_{2n}$.
Using the equation (\ref{As}) one can express the function
$u(\a'_1,\ldots,\a'_{n-1}|\b_{i_1},\ldots,\b_{i_n}|\b_{j_1},\ldots,\b_{j_n})$
for an arbitrary partition as a linear combination of the 
determinants $\text{det}|s_{j_k}(\a_l)|_{1\le k,l\le n-1}$ 
with \\
$j_k=-(n-1),\ldots,-1,
1,\ldots,(n-1)$. One can show that due to the relation (\ref{ss})
only $ \binom{2n}{n}-\binom{2n}{n-1}$ of such determinants are linearly 
independent. 

Thus two requirements (\ref{singlet}) and (\ref{intersect}) together
provide us with sufficient number of linear equations, namely,
$ \binom{2n}{n}$ which, in principle, may be solved and their
solution is unique.

From the equations (\ref{intersect}) we can deduce recurrent relations for
\newline
$v(\a _1,\cdots \a _{n-1}|
\b _{1},\cdots ,\b _{n}|\b _{n+1},\cdots ,\b _{2n})$
which look as follows:
\begin{align}
&v(\a _1,\cdots \a _{n-2},\b|
\b _{1},\cdots ,\b _{n-2},\b-\frac{\pi i}2 ,\b +\frac{\pi i}2
|\b _{n-1},\cdots ,\b _{2n-2})\ =\ \text{exact form}\non\\
&v(\a _1,\cdots \a _{n-2},\b|\b _{1},\cdots ,\b _{n},
|\b _{n+1},\cdots ,\b _{2n-2},\b-\frac{\pi i}2 ,\b +\frac{\pi i}2)
\ =\text{exact form}\non\\
&v(\a _1,\cdots \a _{n-2},\b|\b _{1},\cdots ,\b _{n-1},\b -\frac{\pi i}2
|\b _{n},\cdots ,\b _{2n-2},\b +\frac{\pi i} 2)=\text{exact form}\non\\
&v(\a _1,\cdots \a _{n-2},\b|\b _{1},\cdots ,\b _{n-1},\b+\frac{\pi i}2
|\b _{n},\cdots ,\b _{2n-2},\b-\frac{\pi i}2 )=\non\\&=
\prod\limits _{j=1}^{n-2}(\a _j-\b)\prod\limits _{k=1}^{n-1}
\(\b -\b _k -\frac {\pi i}2\)\(\b -\b _{n+k-1} 
+\frac {\pi i}2\)
\non\\&\times
v(\a _1,\cdots \a _{n-2}|\b _{1},\cdots ,\b _{n-1}
|\b _{n},\cdots ,\b _{2n-2} )
+\text{exact form}
\label{recv}
\end{align}
where "exact form" stands for an expression of the kind:
\begin{align}
&\prod\limits_{j=1}^{n-2} 
\(\a _j-\b\)\sum\limits _{j=1}^{n-2} (-1)^j 
E(\a _j)
m(\a _1,\cdots ,\widehat {\a _j},\cdots ,\a _{n-2}),\non\\
&E(\a )=\prod\limits _{j=1}^{2n-2} \(\a -\b_k+\frac {\pi i}2\)
-\prod\limits _{j=1}^{2n-2} \(\a -\b_k-\frac {\pi i}2\)
\non
\end{align}
for some  skew-symmetric polynomial
of $n-2$ variables $m$.
The relations (\ref{recv}) look undetermined because of presence
of unknown "exact forms" in the RHS. However, knowing 
{\it a priori} degrees of polynomials one can show that this system
defines $v$ completely being in fact even over-determined. Still
we have no key for solving this system. At this point the idea arises
of trying to find the polynomials $v$ using original Jimbo-Miwa formula.
Let us explain this point.

The polynomials $v$ can be decomposed with respect to $s_a$:
\begin{align}
&v(\a _1,\cdots, \a _{n-1})=
\sum\limits _{-(n-1)\le j_1<\cdots < j_{n-1}\le n-1}
K_{j_1,\cdots ,j_{n-1}}
\text{det}\left| s_{j_p}(\a _q)\right|_{p,q=1,\cdots , n-1}
\label{K1}
\end{align}
where irrelevant arguments are omitted. Let us look for 
$K_{j_1,\cdots ,j_{n-1}}$ in the following form:
\begin{align}
&K_{j_1,\cdots ,j_{n-1}}=
\prod_{p=1}^{n-1} \text{sgn}(j_p)\mathcal{L}(s_{-j_1},
\cdots , s_{-j_{n-1}})\label{K2}
\end{align}
where $\mathcal{L}$ is a skew-symmetric 
multi-linear functional. This functional
has to satisfy certain requirements. Recall that it depends upon
$\b _j$ as parameters,
we shall take into account this dependence writing
$$K_{j_1,\cdots ,j_{n-1}}(\b _1,\cdots ,\b _n|\b _{n+1},\cdots , \b _{2n})$$

Our main concern is
the recurrence relations (\ref{recv}).
Now, according to (\ref{recv}), we want to consider the case
$\b _i =\b +\frac {\pi i}2$, $\b _j =\b -\frac {\pi i}2$ for different
choices of $i$ and $j$. Notice that the formula for $c(\a _1,\a _2)$
is symmetric with respect to $\b _j$. So, whatever is the choice of $\b _i$
and $\b _j$ we have:
\begin{align}
\left. c^{(n)}(\a _1,\a _2)
\right| _{\b _i =\b +\frac {\pi i}2, \b _j =\b -\frac {\pi i}2}=
(\a _1-\b)(\a _2-\b)c^{(n-1)}(\a _1,\a _2)+
(\a _1 -\b )E(\a _1)-
(\a _2 -\b )E(\a _2)
\non
\end{align}
where we wrote down $c^{(n)}$ for 
the bilinear form depending on $2n$ variables
$\b _k$ while $c^{(n-1)}$ is similar form depending on 2n-2 $\b$'s
with $\b _i$ and $\b _j$ omitted.
So, for any choice of $\b _i$, $\b _j$ there is a change of basis 
$$ \widehat{s}_a(\a )=C_{a b} \ s_b(\a )$$
such that $C\in Sp(2n-2)$ and
\begin{align}
&\left. \widehat{s}_a(\a )
\right| _{\b _i =\b +\frac {\pi i}2, \b _j =\b -\frac {\pi i}2}=
(\a -\b) s^{(n-1)} _a(\a ),\quad a=-(n-2),\cdots,n-2\non\\
& \left. \widehat{s}_{-(n-1)}(\a )
\right| _{\b _i =\b +\frac {\pi i}2, \b _j =\b -\frac {\pi i}2}=1,\non\\
&\left. \widehat{s}_{n-1}(\a )
\right| _{\b _i =\b +\frac {\pi i}2, \b _j =\b -\frac {\pi i}2}=
(\a -\b)E(\a )
\non
\end{align}
In the formulae (\ref{K1}, \ref{K2}) we can change simultaneously
all $s_a$ to $\widehat{s}_a$. The functional $\mathcal{L}$
depends on $\b _j$ as on parameters. The recurrence
relations (\ref{recv}) are equivalent to the following
ones:
\begin{align}
&\mathcal{L}(\widehat{s}_{a_1},
\cdots ,\widehat{s}_{a _{n-2}}, \widehat{s}_{n-1})(
\b _{1},\cdots ,\b _{n-2},\b-\frac{\pi i}2 ,\b +\frac{\pi i}2
|\b _{n-1},\cdots ,\b _{2n-2})\ =0\label{recv1}\\
&\mathcal{L}(\widehat{s}_{a_1},
\cdots ,\widehat{s}_{a _{n-2}}, \widehat{s}_{n-1})
(\b _{1},\cdots ,\b _{n},
|\b _{n+1},\cdots ,\b _{2n-2},\b-\frac{\pi i}2 ,\b +\frac{\pi i}2)
\ =0\non\\
&\mathcal{L}(\widehat{s}_{a_1},
\cdots ,\widehat{s}_{a _{n-2}}, \widehat{s}_{n-1})
(\b _{1},\cdots ,\b _{n-1},\b -\frac{\pi i}2
|\b _{n},\cdots ,\b _{2n-2},\b +\frac{\pi i} 2)=0\non\\
&\mathcal{L}(\widehat{s}_{a_1},
\cdots ,\widehat{s}_{a _{n-2}}, 
\widehat{s}_{n-1})(\b _{1},\cdots ,\b _{n-1},\b+\frac{\pi i}2
|\b _{n},\cdots ,\b _{2n-2},\b-\frac{\pi i}2 )=\non\\&=
\prod\limits _{k=1}^{n-1}
\(\b -\b _k -\frac {\pi i}2\)\(\b -\b _{n+k-1} 
+\frac {\pi i}2\)
\mathcal{L}(s^{(n-1)}_{a_1},
\cdots ,s^{(n-1)}_{a _{n-2}})(\b _{1},\cdots ,\b _{n-1}
|\b _{n},\cdots ,\b _{2n-2} )
\non
\end{align}
where $a_p> -(n-1)$ $\forall p$, the polynomials $\widehat{s}_a$
are constructed according to above procedure.

In fact the original Jimbo-Miwa formula satisfies some kind of similar
recurrence relations. We cannot go into much details at this point, but
careful development of this idea leads to the following formula for $v$.

An important ingredient of  the formula (\ref{theor}) is the product
$\prod_{j=1}^{2n}\varphi(\s-\b_j)$. Asymptotically as $\s\to\infty$ one has:
$$\prod_{j=1}^{2n}\varphi(\s-\b_j)\simeq e^{-n\s+\frac 1 2\sum\b_j}\Phi(\s)$$
where $\Phi(\s)$ are 
asymptotic series with the following properties:
\begin{align}
\Phi (\s)=\s ^{-n}(1+c_1(\b)\s^{-1}+c_2(\b)\s^{-2}+\cdots)
\label{Phiexp}
\end{align}
\begin{align}
&\Phi(\s+2\pi i)=\Phi(\a)\frac{P(\s+\frac{\pi i} 2)}{P(\s+\frac{3\pi i} 2)}
\non\\
&\Phi(\s)\Phi(\s+\pi i) = \frac{1}{P(\s+\frac{\pi i}{2})}
\label{Phicond}
\end{align}
where 
\begin{align}
P(\s)=\prod_{j=1}^{2n} (\s-\b_j)
\label{Ps}
\end{align}
Consider polynomials $p_1,\cdots ,p_{n-1}$ which are taken as
a subset of $s_a$. Using the properties of the function $\Phi$
one can check that for such a polynomials
\begin{align}
\text{res}_{\infty}(p_{i}(\s)\Phi(\s))=0
\label{res0}
\end{align}
Then for every $p_i$ let us define:
\begin{align}
X_{i}(\s)=\Delta^{-1}(p_{i}(\s)\Phi(\s))
\label{X}
\end{align}
where by the definition 
\begin{align}
(\Delta f)(\s) = f(\s+\pi i) - f(\s - \pi i)
\label{Delta}
\end{align}
The r.h.s. of eq. (\ref{X})
is well defined as asymptotic series since $p_i\Phi$ does
not have residue (\ref{res0}). Our first idea 
for definition of $\mathcal{L}(p_1,\cdots ,p_{n-1})$
was as follows:
\begin{align}
&\mathcal{L}^{0}(p_1,\cdots ,p_{n-1})=
\int\limits _{\Gamma _1}\frac{d\sigma _1}{2\pi i}
\cdots\int\limits _{\Gamma _{n-1}}\frac {d\sigma _{n-1}}{2\pi i}
\ h_0(\s _1,\cdots ,\s _{n-1})
\prod\limits _{j=1}^{n-1}X_j(\s _j)\Phi (\s _j)\non
\end{align}
where
\begin{align}
&h_0(\s _1,\cdots ,\s _{n-1})=
\sum_{\pi\in S_{n-1}} (-1)^{
\text{sgn}
\text{$\pi$}
}
D(\s _{\pi\{ 1\}},\cdots \s _{\pi\{n-1\}}|
\b _1,\cdots ,\b _{2n})_{-\cdots - +\cdots  +}
\label{F0}
\end{align}
and the function $D$ is given by the formula (5) of the paper \cite{bks1}
\begin{align}
&D(\a _1,\cdots \a _{n-1}|\b _1,\cdots ,\b _{2n})_{-\cdots - +\cdots +}=
\non\\
&=\prod\limits _{r>s}\frac 1 {\a _r-\a _s-\pi i}
\prod\limits _{k}\prod\limits _{j>k}\(\a _k -\b _j+\frac {\pi i} 2\)
\prod\limits _{j<k}\(\a _k -\b _j-\frac {\pi i} 2\)
\non\\
&\times \sum\limits _{l=1}^{n} \(2\sum \a _k+\b _l+\pi i(2l-1)\)
\prod\limits _{j\ge l}
\frac {\a _j -\b _j-\frac {\pi i} 2}{\a _j-\b_{j+1}+\frac {\pi i} 2}
\label{D}
\end{align}
Due to the properties of $D$ which ensure correct form of residues 
in Jimbo-Miwa \cite{JM} formulae one can expect that the formula 
(\ref{F0}) satisfies correct recurrence relations. It is almost the case, but
some corrections are needed. The trouble is that $h_0$ has poles
at $\s _i=\s _j \pm \pi i$. So, first of all we have to explain how
the contours $\Gamma _j$ are drawn. Let us use the following prescription:

\vskip 1cm
\hskip 2cm  
\epsffile{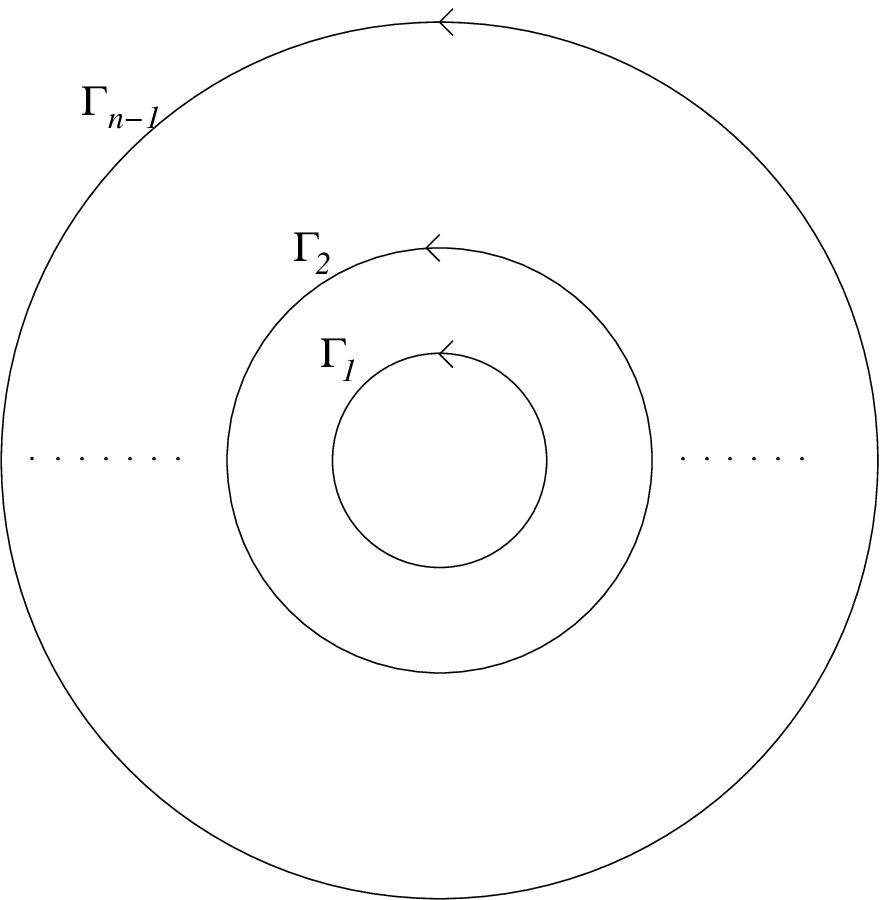}
\vskip 0.2cm
\noindent

Namely, we require that for any $j$ $|\s_j|\ll |\s_{j+1}|$ where
$\s _j\in\Gamma_j$ and $\s_{j+1}\in\Gamma_{j+1}$. This definition,
however brakes the skew-symmetry with respect to polynomials
$p _i$ which is crucial for us. Necessary improvement is done as follows.
Let us introduce some notations.
Let $U$ be a set  $U=\{1,\ldots,n-1\}$ which is decomposed
into the union of three subsets
$$
U = S \bigcup \overline{ S} \bigcup T
$$
where $S$ 
and $\overline{ S}$ are the subsets of $ U$ of the length $p$ with
$0\le p\le [\frac{n-1}{2}]$ and $ T$ is their complementary subset
of the length $n-1-2p$
\begin{align}
&S = \{i_1<i_2<\ldots <i_p\}\nonumber\\
&\overline{ S} = \{\bar{ i_1},\;\bar{ i_2},\;\ldots,\;\bar{  i_p}\}\quad\quad
i_1<\bar{  i_1},\;i_2<\bar{  i_2},\ldots,i_p<\bar{  i_p}\nonumber\\
& T = \{j_1<j_2<\ldots <j_{n-1-2p}\}
\label{SST}
\end{align}
Let us note that there is the ordering in the subset $S$ while
for the subset $\overline{ S}$ the ordering is not required.

Then let us define an object
\begin{align}
&I_{S,\overline{ S},T} \equiv 
\int_{\Gamma_{i_1}}\frac{d\s_{i_1}}{2\pi i}
\ldots
\int_{\Gamma_{i_p}}\frac{d\s_{i_p}}{2\pi i}
\int_{\Gamma_{j_1}}\frac{d\s_{j_1}}{2\pi i}
\ldots
\int_{\Gamma_{j_{n-1-2p}}}\frac{d\s_{j_{n-1-2p}}}{2\pi i}
\non\\
&\prod_{q=1}^p X_{i_q}(\s_{i_q})p_{\bar{ i_q}}(\s_{i_q})\Phi(\s_{i_q})
\prod_{q=1}^{n-1-2p} X_{j_q}(\s_{j_q})\Phi(\s_{j_q})
h_p(\s_{i_1},\ldots,\s_{i_p}|\s_{j_1},\ldots,\s_{j_{n-1-2p}})
\label{I}
\end{align}
where the functions $h_p$ are obtained from $h_0$ by a procedure
which will be described later.
\begin{align}
&\mathcal{L}(p_1,\cdots ,p_{n-1})=
\sum\limits _
{ U = S 
\bigcup \overline{ S }\bigcup  T}\epsilon (S, \overline{ S}, T)
I_{S,\overline{ S},T}\label{fin}
\end{align}
where $\epsilon (S, \overline{ S}, T)$  is the sign of the permutation $\pi$: 
$$
\pi: \quad (1,\ldots,n-1)\rightarrow (i_1,\overline i_1,i_2,\overline 
i_2,\ldots,
i_p,\overline i_p,j_1,\ldots,j_{n-1-2p})
$$

Let us return to $h_p$. Introduce the following notations:\newline
$\delta$ for the finite difference:
$$\delta (f)(\sigma)=f(\sigma +\pi i)-f(\sigma),$$
$R_k$ for normalized residue
$$
R_k=\frac 1 {P(\s_k+\frac{\pi i}{2})} 
\text{res}_{\s_{k+1}=\s_k+\pi i}
$$
where $P(\s)$ is given by (\ref{Ps})
and $D_l$ is the ``exact form'' taken in the variable $\s_l$
\begin{align}
D_lf(\ldots|\s_l|\ldots)=
&\frac{f(\ldots|\s_l|\ldots)}{P(\s_l+\frac{3\pi i}{2})}-
\frac{f(\ldots|\s_l-2\pi i|\ldots)}{P(\s_l-\frac{3\pi i}{2})}&
\label{D_l}
\end{align}

Then
\begin{align}
&h_p(\s_1,\s_3,\ldots,\s_{2p-1}|\s_{2p+1},\s_{2p+2},\ldots,\s_{n-1})=
\prod_{k=1}^{p}\d^{-1}_{\s_{2k-1}}
\biggl[
\prod_{k=1}^{p}R_{2k-1}\; h_0(\s_1,\s_2,\ldots,\s_{n-1})
\non\\
&
-\sum_{r=1}^{
\text{min}
(p,[\frac{n-2}{2}])}(-1)^r
\sum_{1\le m_1<\ldots<m_r\le p}
\prod_{
\begin{tabular}{c}
\small{$1\le k\le p$}\\
\small{$k\ne\{m_1,\ldots,m_r\}$}\\
\end{tabular}
}
R_{2k-1}
\sum_{
\begin{tabular}{c}
\small{$1\le l\le n-1$}\\
\small{$l\ne\{2m_1-1,2m_1,\ldots,2m_r-1,2m_r\}$}\\
\end{tabular}
}
\non\\&
(-1)^lD_l f_r(\s_{2m_1-1},\ldots,\s_{2m_r-1}|\s_l|\s_1,\ldots,
\widehat{\s_{2m_1-1}},\widehat{\s_{2m_1}},\ldots,\widehat{\s_l},\ldots,
\widehat{\s_{2m_r-1}},\widehat{\s_{2m_r}},\ldots,\s_{n-1})\biggr]
\label{hp}
\end{align}
The functions $f_r$ are chosen in such a way that 
\newline
1. Taking "primitive" 's $\delta ^{-1}$ is possible.
\newline
2. $
h_p(\s_1,\s_3,\ldots,\s_{2p-1}|\s_{2p+1},\s_{2p+2},\ldots,\s_{n-1})$ 
is symmetric with respect to \newline $\s_1,\s_3,\ldots,\s_{2p-1}$ and
skew-symmetric with respect to 
$\s_{2p+1},\s_{2p+2},\ldots,\s_{n-1}$ or equivalently\\
$
f_r(\s_1,\s_3,\ldots,\s_{2r-1}|\s_{2r}|\s_{2r+1},\s_{2r+2},\ldots,
\s_{n-2r-2})$ is symmetric w.r.t. $\s_1,\s_3,\ldots,\s_{2r-1}$
and skew-symmetric w.r.t. 
$\s_{2r+1},\s_{2r+2},\ldots,\s_{n-2r-2}$ 
\newline
3. For the pair of variables $\s_{2r},\s_{2r+1}$ the only non-zero
residues are at the points\\
 $\s_{2r+1}=\s_{2r}+3\pi i$,
$\s_{2r+1}=\s_{2r}+\pi i$ and $\s_{2r+1}=\s_{2r}-\pi i$ and 
\begin{align}
&\text{res}_{\s_{2r+1}=\s_{2r}+3\pi i} 
f_r(\s_1,\s_3,\ldots,\s_{2r-1}|\s_{2r}|\s_{2r+1},\s_{2r+2},\ldots,
\s_{n-2r-2})=\non\\
&\text{res}_{\s_{2r+1}=\s_{2r}} 
f_r(\s_1,\s_3,\ldots,\s_{2r-1}|\s_{2r}+\pi i|\s_{2r+1},\s_{2r+2},\ldots,
\s_{n-2r-2})
\label{condf}
\end{align}
\newline
Some additional requirements on these functions $f_r$ must be
satisfied. For the moment we were not able to construct
these functions completely for arbitrary $n$, formulae
for $n=2,3,4,5,6$ are given in Appendix. 

Notice two remarkable properties of the formulae 
(\ref{I} , \ref{fin}). First, in order to achieve skew-symmetry
we need to add some integral with coupled $p_i$ and $p_{\bar{i}}$. 
But we do not need to consider more complicated
integrals in which three or more polynomials are bounded
together. Second, due to the fact that $p_i\circ p_j=0$
the ambiguity in definition of $\delta ^{-1}$ in the formula for
$h_p$ is irrelevant. The definition of the pairing $\circ$ is
defined by the formula (15) of the paper \cite{bks1}.

\section{The correlation functions}

Let us remind the reader the formula from \cite{JM}
which relates a solution of the qKZ on level -4 to
any correlation function of the XXX model:
\begin{align}
\langle \prod^n_{j=1} E^{\e_j,\e'_j} \rangle =
{g(\b_1,\ldots,\b_n,\b_n+\pi i,\ldots,\b_1+\pi i)}^
{-\e'_1,\ldots,-\e'_n,\e_n,\ldots,\e_1}
\label{cor1}
\end{align}
Let us also note that this solution should also satisfy 
some additional constraint, namely, that
\begin{align}
&{g(\ldots,\b_{j-1},\b_j,\b_j-\pi i,\b_{j+2},\ldots)}^
{\ldots,\e_{j-1},\e_j,\e_{j+1},\e_{j+2},\ldots}=\non\\&=
\d^{\e_j,-\e_{j+1}}\;
{g(\ldots,\b_{j-1},\b_{j+2},\ldots)}^
{\ldots,\e_{j-1},\e_{j+2},\ldots}
\label{cond3}
\end{align}
We had complete set of singlet solutions to qKZ labeled by the multiindex
$\{k_1,\cdots ,k_{n-1}\}$. The solution satisfying (\ref{cond3}) is the one
corresponding to $k_j=n-1-2j$. This solution is divisible by $\sum e^{\b _k}$.
So, in what follows we consider the function:
\begin{align}
&g(\b _1,\cdots ,\b _{2n})=
\frac 1
{\sum e^{\b _j}}\ g_{\{-(n-1),-(n-3),\cdots , (n-3)\}}(\b _1,\cdots ,\b _{2n})
\end{align}
Rewrite the components of $g(\b _1,\cdots ,\b _{2n})$ as follows
\begin{align}
&g(\b _1,\cdots ,\b _{2n})^{\e _1,\cdots ,\e _{2n}}=
\frac {e^{\half\sum \b _j}}
{\sum e^{\b _j}}
\prod\limits_ {i<j}\frac{1}{\zeta (\b _i-\b _j)}
\non\\ &\times
\sum\limits_{-(n-1)\le j_1< j_2<\cdots <j_{n-1}\le n-1}
P^{\e _1,\cdots ,\e _{2n}}_{j_1, \cdots, j_{n-1}}(\b _1,\cdots ,\b _{2n})
\text{det}\(I_{j_l,m}\)_{l,m=1,\cdots ,n-1}\label{sol}
\end{align}
where polynomials
\begin{align}
&P^{\e _1,\cdots ,\e _{2n}}_{j_1, \cdots, j_{n-1}}
(\b _1,\cdots ,\b _{2n})=
\sum\limits
_{\{1,\cdots,2n\}=\{k_1,\cdots ,k_n\}\cup \{l_{1},\cdots ,l_{n}\}}
K_{j_1,\cdots,j_{n-1}}
(\b _{k_1},\cdots ,\b _{k_n}|\b _{l_1},\cdots ,\b _{l_n})
\non\\
&
\times\prod
\limits _{p,q=1}^n\frac {\b _{l_q}-\b _{k_p}+\pi i} 
{\b _{l_q}-\b _{k_p}}
{w^{\dag}}_{\e'_1,\cdots,\e'_{2n}}^{\e_1,\cdots,\e_{2n}}
(\b_1,\cdots ,\b _{2n})
\label{PvK}
\end{align}
and as above $\e'_{k_p}=-$ and $\e'_{l_q}=+$.
Here we have introduced  the integrals:
\begin{align}
I_{j ,m}=\int\limits _{-\infty}^{\infty}d\a\prod\limits
_{p=1}^{2n}    \varphi (\a -\b _p)
s_j(\a)e^{(n-1-2m)\a}
\non
\end{align}
where the functions $\varphi$ and $\zeta$ are given by the 
formulae (\ref{phi}) and (\ref{zeta}) respectively. 
We want to split $\b$'a into pairs:
\begin{align}
\b_k=\la _k-\frac{\pi i}{2}+i\delta _k,\quad
\b_{2n-k+1}=\la _k+\frac{\pi i}{2}-i\delta _k 
\label{limdelta}
\end{align}
and to take limit $\delta _k\to 0$.
It is very convenient to keep $\delta _k$ different.

The function $\varphi (\a )$ satisfies the equation:
$$\varphi (\a+\frac {\pi i}2 )\varphi (\a -\frac {\pi i}2)=
\frac {4\pi^3} {\a\sinh(\a)}$$
In the limit $\delta _k\to 0$ the contour of integration is pinched
by two poles which may give a singularity. Let us study the leading
behaviour. For $s_{-k}$ the situation is simple:
\begin{align}
&I_{-k ,m}=\pi (4\pi )^n\sum\limits _{p=1}^n \frac 1 {\delta _p}
\prod\limits _{q\ne p}\frac 1 {(\la _p-\la _q)\sinh (\la _p-\la _q)}
e^{(n-1-2m)\la _p}\la _p^{n-1-k}+\mathcal{O}(1)\non
\end{align}
so, the leading part of the integrals is expressed in terms
of elementary functions. 

For $s_k$ with $k>0$ the situation is more complicated.
It is more convenient to use here the explicit formula
for the polynomials 
$$
s_k(\a)\equiv A_k(\a|\b_1,\cdots,\b_n|\b_{n+1},\cdots,\b_{2n})
$$
which follows form the formula (\ref{gener}) for the generating
function
\begin{align}
&A_k(\a|\b_1,\cdots,\b_n|\b_{n+1},\cdots,\b_{2n})=\non\\
&\prod_{j=1}^n(\a-\b_j-\frac{\pi i}{2})
Q_k(\a-\frac{\pi i}{2}|\b_{n+1}-\frac{\pi i}{2},\cdots,\b_{2n}-\frac{\pi i}{2})
+\non\\
+
&\prod_{j=1}^n(\a-\b_{n+j}+\frac{\pi i}{2})
Q_k(\a+\frac{\pi i}{2}|\b_{1}+\frac{\pi i}{2},\cdots,\b_{n}+\frac{\pi i}{2})
\label{Ak}
\end{align}
with the polynomials
\begin{align}
Q_k(\a|\l_1,\cdots,\l_n)=\sum_{l=1}^k
\biggl((\a+\frac{\pi i}{2})^l-(\a-\frac{\pi i}{2})^l\biggr)
(-1)^{k-l}
\s_{k-l}(\l_1,\cdots,\l_n)
\label{Qk}
\end{align}
and the standard symmetric functions of $n$ variables 
$\s_k(\l_1,\cdots,\l_n)$ defined as follows
$$
\prod_{j=1}^n(\l+\l_j)=\sum_{l=0}^n\l^l\s_{k-l}(\l_1,\ldots,\l_n)
$$

Actually, the multipliers $\prod (\a -\b _j\pm\frac {\pi i}2)$
in (\ref{Ak})
cancel singularities of $\varphi (\a -\b _j)$ closest to the real axis.
So, the pinching does not take place and, hence, the result is finite.
Denote:
\begin{align}
&I_{k,m}=I_{k,m}^0+\sum i\delta _p I_{k,m}^p+\mathcal{O}(\delta ^2)\non
\end{align}
For $I_{k,m}^0$ we have:
\begin{align}
I_{k,m}^0=
&\int\limits _{\mathbb{R}+i0}d\a
\prod_{j=1}^n\frac{4\pi^3}{ \sinh(\a -\la_j)}
e^{(n-1-2m)\a}Q_k(\a -\frac {\pi i}2|\la _1, \cdots, \la _n)+
\non\\+
&\int\limits _{\mathbb{R}-i0}d\a
\prod_{j=1}^n\frac{4\pi^3}{ \sinh(\a -\la_j)}
e^{(n-1-2m)\a}Q_k(\a +\frac {\pi i}2|\la _1, \cdots, \la _n)=0
\non
\end{align}
So, the leading contribution is of the first order
in $\delta$'s. Consider $I_{k,m}^p$:
\begin{align}
&I_{k,m}^p=\non\\
&\int\limits _{\mathbb{R}+i0}d\a
\prod_{j=1}^n\frac{4\pi^3} {\sinh(\a -\la_j)}
e^{(n-1-2m)\a}
\(\chi (\a-\la _p)-\frac 1 {\a-\la _p}-\frac{\partial} {\partial \la _p}\)
Q_k(\a -\frac {\pi i}2|\la _1, \cdots, \la _n)
\ +\non\\+
&\int\limits _{\mathbb{R}-i0}d\a\prod_{j=1}^n\frac{4\pi^3} {\sinh(\a -\la_j)}
e^{(n-1-2m)\a}
\(\chi (\a-\la _p)+\frac 1 {\a-\la _p}+\frac{\partial} {\partial \la _p}\)
Q_k(\a +\frac {\pi i}2|\la _1, \cdots, \la _n)
\ 
\non
\end{align}
where
\begin{align}
&\chi (\a)=\frac d {d\a}\(\log\frac{\varphi (\a -\frac{\pi i} 2)}
{\varphi (\a +\frac{\pi i} 2)}\)=2\pi i\sum\limits_{k=1}^{\infty}
(-1)^k\frac{k}{\a ^2+(\pi k)^2}
\non
\end{align}

Using the definition (\ref{Qk}) of $Q_k$ we present
$I_{k,m}^p$ as follows:
$$I_{k,m}^p=\sum\limits _{l=1}^k
(-1)^{k-l}\sigma _{k-l}(\la _1, \cdots ,\la _n)J_{l,m}^{p}$$
Then
\begin{align}
&J_{l,m}^p=\non\\
&\int\limits _{\mathbb{R}+i0}d\a
\prod_{j=1}^n\frac{4\pi^3} {\sinh(\a -\la_j)}
e^{(n-1-2m)\a}
\(
\chi (\a-\la _p)
\((\a-\pi i)^l-\a^l\)
+\pi i\frac {(\a-\pi i)^l-\la _p^l}{(\a-\la _p)(\a-\la _p-\pi i)}\)
\ +\non\\+
&\int\limits _{\mathbb{R}-i0}d\a
\prod_{j=1}^n\frac{4\pi^3} {\sinh(\a -\la_j)}
e^{(n-1-2m)\a}
\(\chi (\a-\la _p)
\(\a^l-(\a+\pi i)^l\)
-\pi i\frac {(\a+\pi i)^l-\la _p^l}{(\a-\la _p)(\a-\la _p+\pi i)}
\)
\ 
\non
\end{align}
Here we have used the following identity
$$
\frac{\partial}{\partial\l_p}\s_k(\l_1,\cdots,\l_n)=
\sum_{l=0}^{k-1}(-\l_p)^{k-l-1}\s_l(\l_1,\ldots,\l_n)
$$

Let us denote
$$f(\a)=\a ^l-\la _p^l$$
then using the identities:
\begin{align}
&\chi (\a +\pi i)+\chi(\a)=\frac 1 {\a +\pi i}-\frac 1 {\a }
\\
&
\chi (\a)+\chi(\a-\pi i)=
\frac 1 {\a }-\frac 1 {\a -\pi i}
\non
\end{align}
one finds:
\begin{align}
J_{l,m}^p=&\int\limits _{\mathbb{R}+i0}d\a
\prod_{j=1}^n\frac{4\pi^3} {\sinh(\a -\la_j)}
e^{(n-1-2m)\a}
\(f(\a-\pi i)
\chi (\a-\la _p-\pi i)+f(\a)
\chi (\a-\la _p)\)-\non\\
-&\int\limits _{\mathbb{R}-i0}d\a
\prod_{j=1}^n\frac{4\pi^3} {\sinh(\a -\la_j)}
e^{(n-1-2m)\a}
\(f(\a+\pi i)
\chi (\a-\la _p+\pi i)+f(\a)
\chi (\a-\la _p)\)=\non\\&=
2\(\int\limits _{\mathbb{R}+i0}-\int\limits _{\mathbb{R}-i0}\)d\a
\prod_{j=1}^n\frac{4\pi^3} {\sinh(\a -\la_j)}
e^{(n-1-2m)\a}\chi (\a-\la _p)\(\a ^l-\la_p^l\)=\non\\&=
4\pi i(4\pi^3)^n \sum\limits _{q\ne p}
\prod\limits _{j\ne q}\prod\frac{4\pi^3} {\sinh(\a -\la_j)}
e^{(n-1-2m)\la _q}\chi(\la_p-\la _q)
(\la _p^l-\la _q^l)\non
\end{align}

Returning to $I_{k,m}^p$:
\begin{align}
I_{k,m}^p&=4\pi i (4\pi^3)^n(-1)^k\sum\limits _{q\ne p}
\prod\limits _{j\ne q}
\frac 1 {\sinh (\la _q-\la _j)}e^{(n-1-2m)\la _q}\chi(\la_p-\la _q)
\non\\ &\times
(\sigma _k(\la _1,\cdots ,\widehat{\la _p},\cdots ,\la _n)-
\sigma _k(\la _1,\cdots ,\widehat{\la _q},\cdots ,\la _n))
\non
\end{align}
Let $\tilde{I}_{j,m}$ ($|j|=1,\cdots ,n-1$) correspond to
a leading term of $I_{j,m}$ in $\delta$'s.
It can be rewritten in the following form:
\begin{align}
&\tilde{I}_{j,m}=\sum\limits _{p=1}^n A_{j,p}B_{p,m}\non
\end{align}
where $A$ and $B$ are, respectively, $2(n-1)\times n$ and $n\times (n-1)$
matrices with the following matrix elements:
\begin{align}
&A_{-k,p}=\pi(4\pi^3)^n
\prod\limits _{q\ne p}\frac 1 {\la _p-\la_q}\ \la _p^{n-1-k},\non\\
&A_{k,p}=(-1)^{k-1}4\pi(4\pi^3)^n
\sum\limits _{q\ne p} \delta_p\delta_q \chi(\la _p-\la _q)
(\sigma _k(\la _1,\cdots ,\widehat{\la _p},\cdots ,\la _n)-
\sigma _k(\la _1,\cdots ,\widehat{\la _q},\cdots ,\la _n))
\non\\
&B_{p,m}=\frac 1{\delta _p}\ 
\prod_{j\ne p}\frac 1 {\sinh (\la _p-\la _j)}e^{(n-1-2m)\la _p}
\non \end{align}
Notice the following simple property:
\begin{align}
&\sum\limits _{p=1}^n\ A_{j,p}=0
\label{A=0}
\end{align}
Take a partition $\{j_1,\cdots ,j_{n-1}\}$ and introduce
the following object
\begin{align}
&A_{j_1,\cdots ,j_l}\equiv\text{det}(A_{j_l,r})|_{r\ne p}
\non
\end{align}
Then the identity
(\ref{A=0}) implies that this definition does not depend
on the choice of $p$. 
There is one more identity which replaces Riemann bilinear
relation:
\begin{align}
&\sum_{k=1}^{n-1}\(A_{-k,p}A_{k,q}-A_{-k,q}A_{k,p}\)=0\label{rbi}
\end{align}

Calculating minors of the
matrix $B$ one finds:
\begin{align}
&\text{det}(\tilde{I}_{j_l,m})|_{l,m =1,\cdots n-1}=
2^{\frac{(n-1)(n-2)}{2}}
A_{j_1,\cdots ,j_l}\prod _{p=1}^n\frac 1 {\delta _p}
\frac{\sum(\delta_p e^{\la _p})\exp (-\sum \la _p)}
{\prod\limits _{i<j}\sinh (\la _i-\la _j)}\label{det}
\end{align}
Let us return to the formula (\ref{sol}). It can be shown that
$$\zeta(\la)\zeta (-\la)\zeta (\la-\pi i)\zeta (-\la -\pi i)=
C^2\frac {\la} {\sinh (\la ) }$$
where the constant
$$
C=\frac{\zeta(\frac{\pi i}2)\zeta(-\frac{\pi i}2)}
{\Gamma(\frac 1{4})^2(\frac{\pi}2)^2}
$$
So, the formula (\ref{sol}) can be rewritten as follows:
\begin{align}
&g(\la_1 -\frac {\pi i} 2+i\delta _1,\cdots ,\la _{1}+
\frac {\pi i} 2-i\delta _1)
^{\e _1,\cdots ,\e _{2n}}=\biggl[\prod _{p=1}^n\frac 1 {\delta _p}\biggr]
\frac{2^{n(n-3)/2}}{C^{n(n-1)}}\prod_{i<j}\frac{1}{\l_i-\l_j}
\non\\&\times
\sum\limits_{-(n-1)\le j_1< j_2<\cdots <j_{n-1}\le n-1}
P^{\e _1,\cdots ,\e _{2n}}_{j_1, \cdots, j_{n-1}}
(\la_1 -\frac {\pi i} 2+i\delta _1,\cdots ,
\la_1 +\frac {\pi i} 2-i\delta _1)
A_{j_1,\cdots ,j_l}\(1+\mathcal{O}(\delta)\)\label{sol1}
\end{align}
{\it A priori} we know that the result of specification
$\b _k=\la _k-\frac {\pi i} 2$, $\b _{2n-k-1}=
\la _k+\frac {\pi i} 2$ is finite.
The only source of possible singularities in $\delta$'s is the multiplier
$\prod\frac 1 {\delta _p}$ in RHS of (\ref{sol1}). The sum is  Tailor
series in $\delta$'s.  Hence, in the sum\newline
1. All the coefficients in front of those terms
which are not divisible by $\prod \delta _p$ must vanish.
\newline
2. In the final result only one term is important: that containing
$\prod \delta _p $. It means, in particular that
the terms divisible by $\delta _p^2$ for some $p$ do not contribute
into result.
 
Recall that $A_{j_1,\cdots ,j_{n-1}}$ is constructed out of
$A_{j,p}$. 
Let us denote the number of positive $j$'s in the multiindex by
$l(\{j\})$.
Obviously
\begin{align}
&A_{\{j\}}=\mathcal{O}(\delta ^{2l(\{j\})})\label{estim}
\end{align}
The second property shows that $A_{\{j\}}$ for which
$$l(\{j\})>\[\frac n 2\]$$
do not
contribute to the final result.
 
Now we need the following \newline
{\bf Assumption.}
If  $l(\{j\})\le\[\frac n 2\]$ then
\begin{align}
&P^{\{\e \}}_{\{ j\}}(\la_1 -\frac {\pi i} 2+i\delta _1,\cdots ,\la_1 +
\frac {\pi i} 2-i\delta _1)=\mathcal{O}(\delta ^{n-2l(\{j\})})\label{estim1}
\end{align}

Unfortunately, for the moment we are not able to prove it.
Let us mention at least some arguments why we believe in this Assumption? 
First, suppose the degree in $\delta $ is smaller
then in (\ref{estim1}). Then some huge cancellation must take place because
all contributions from leading term of asymptotic described by
$A_{\{j\}}$ are singular in the limit. The estimation (\ref{estim1})
is valid for $n=2,3$, these cases we were able to calculate by Maple.
Certainly, the best way to prove the Assumption is to calculate
the polynomials $P^{\{\e \}}_{\{ j\}}$. 
These polynomials are related to the polynomials $K_{j_1,\cdots,j_{n-1}}$
through the formula (\ref{PvK}). The latter polynomials are given
by the formulae (\ref{K2}), (\ref{I}) and (\ref{fin}). The main problem here
is to take the limit $\d_k\rightarrow 0$ for the integrals (\ref{I})
when the spectral parameters $\b_k$ are taken in the form (\ref{limdelta}).
We leave this question for further publication.

Let us suppose that the Assumption is true. Then still important cancellations
must occur because  only one kind of terms of the order $\delta ^n$ must
remain in the Tailor series: that with $\prod\delta _p$. 
The result will look like:
\begin{align}
&g(\la_1 -\frac {\pi i} 2,\cdots ,\la _{1}+\frac {\pi i} 2)
^{\e _1,\cdots ,\e _{2n}}=
\non\\&=
\sum\limits _{m=0}^{\[\frac n 2\]}
\sum\limits _{k_1,\cdots ,k _{2m}}
Q_{k_1,\cdots ,k _{2m}}^{\e _1,\cdots ,\e _{2n}}(\la _1,\cdots, \la _n)
\chi(\la _{k_1}-\la _{k_2})\cdots \chi(\la _{k_{2m-1}}-\la _{k_{2m}})
\label{fin1}
\end{align}
where $Q_{k_1,\cdots ,k _{2m}}^{\e _1,\cdots ,\e _{2n}}(\la _1,\cdots \la _n)$
are rational functions with denominators containing only products
of $\la _i-\la _j$.
The requirement that terms in Tailor series with $\delta _p^2$ must disappear
implies that $k_i\ne k_j$ $\forall i,j$.
We have checked for $n=2$ and $n=3$ that the functions
$Q_{k_1,\cdots ,k _{2m}}^{\e _1,\cdots ,\e _{2n}} $ have
additional multipliers, namely,
$$
Q_{k_1,\cdots ,k _{2m}}^{\e _1,\cdots ,\e _{2n}}(\l_1,\cdots,\l_n) =
\prod_{j=1}^m((\l_{k_{2j-1}}-\l_{k_{2j}})^2+\pi^2)
\tilde A_{k_1,\cdots ,k _{2m}}^{\e_1,\cdots,\e_{2n}}(\l_1,\cdots,\l_n)
$$
where $\tilde A_{k_1,\cdots ,k _{2m}}^{\e_1,\cdots,\e_{2n}}$ are some rational
functions.
Doing a substitution like $\l_k\rightarrow \pi (z_k-\frac{i}{2})$ 
and using the formula (\ref{cor1}) we arrive at the formula 
\begin{align}
\langle \prod^n_{j=1} E^{\e_j,\e'_j} \rangle =
\sum\limits _{m=0}^{\[\frac n 2\]}
\sum\limits _{k_1\ne\cdots \ne k _{2m}}
A_{k_1,\cdots ,k _{2m}}^{-\e' _1,\cdots ,-\e'_n,\e_n,\cdots,\e _1}
(z_1,\cdots, z_n)
G(z_{k_1}-z_{k_2})\cdots G(z_{k_{2m-1}}-z_{k_{2m}})
\label{cor2}
\end{align}
where $A_{k_1,\cdots ,k _{2m}}^{\e _1,\cdots ,\e _{2n}}(z_1,\cdots, z_n)$ 
are rational functions related in an obvious way to the functions \\
$\tilde A_{k_1,\cdots ,k_{2m}}^{\e_1,\cdots,\e_{2n}}(\l_1,\cdots,\l_n)$
and the function $G$ is defined by the formula (3.6) of the work
\cite{bks}.
We see wonderful agreement with the ansatz (3.20) from the above paper
for the emptiness formation probability $P_n(z_1,\cdots,z_n)$
which was based on quite different arguments. 

\section{Conclusion}

In this paper we have applied a new form of solution to the
qKZ on level -4 for the correlation functions of the XXX model.  
They are given by the formula (\ref{cor2}) in a full agreement
with the result of paper \cite{bks}. 
The mathematical reasons for a reducibility of  solutions
of the qKZ on level -4 and the correlation functions to 
one-dimensional integrals is the special kind of the cohomologies
of the deformed Jacobi varieties. 
Our previous conjecture that in the homogeneous limit $z_j\rightarrow 0$
both the emptiness formation probability and other correlation functions
may be expressed in terms of the Riemann zeta function at odd arguments
and rational coefficients \cite{bk1,bk2} follows from the formula
(\ref{cor2}) and the expansion (3.13) of the paper \cite{bks}.

Unfortunately, there are still some technical problems that were not solved
or solved only partially in this paper. 
First we are not completely satisfied with the formulae 
(\ref{K2}), (\ref{I}) and (\ref{fin}) which are necessary for defining
the polynomials $K_{j_1,\cdots,j_{n-1}}$.
In particular, we have not succeeded in finding
a general formula for the functions $f_r$ from the formula (\ref{hp}).
We have done it completely for $n\le 6$ (see Appendix). For a generic case
we have only formulated some general conditions for these functions.
We may hope that, probably, some more explicit formulae should exist 
for those polynomials.
 
Also the challenging problem is to define the rational functions 
$A^{k_1,\cdots ,k _{2m}}_{\e _1,\cdots ,\e _{2n}}(z_1,\cdots, z_n)$
explicitly using the results of the Section 2. 

We hope to do it in our further publication. We also hope to
generalize all the above results to the case of the XXZ spin chain.

\section{Acknowledgements}

The authors would like to thank R. Flume, F.~G{\"o}hmann,   
A.~Kl{\"u}mper  and B.~McCoy
for useful discussions.
This research  has been supported by the following grants:
NSF grant PHY-9988566, the Russian Foundation of Basic Research
under grant \# 01--01--00201, by INTAS under grants \#00-00055 and \# 00-00561.
HEB would like to thank the administration of 
the Max-Planck Institute for Mathematics
for hospitality and perfect conditions for the work.

\section{Appendix}

Here we show how the scheme generally described in the 
previous Section works for the cases $n=2,3,\ldots, 6$.

1. {\bf The case $n=2$}

This case is rather trivial because the function $D$ defined 
in (\ref{D}) does not have any poles and 
$$
h_0(\s) = D(\s|\b_1,\b_2,\b_3,\b_4)_{--++}
$$
The formula (\ref{fin})
contains only one term for $p=0$ which is the one-fold integral:
\begin{align}
&\mathcal{L}(p_1)= \int_{\Gamma_1}\frac{d\s_1}{2\pi i}
X_1(\s_1)\Phi(\s_1) h_0(\s_1)&
\label{n=2}
\end{align}
and the function $X_1$ is defined 
by the formula (\ref{X}) through a polynomial $p_1$ which in this
case is one of the polynomials $s_1$ or $s_{-1}$.

2. {\bf The case $n=3$}

This is the first non-trivial case because one faces two-fold integrals
for which the integration order is essential.

Let us represent the function $D$ from (\ref{D}) in 
the following form:
\begin{align}
&D(\s_1,\s_2|\b_1,\ldots,\b_6)_{---+++} = 
\frac{g(\s_1,\s_2)}{\s_2-\s_1-\pi i}&
\label{D3}
\end{align}
where $g$ is a polynomial for which the dependence on the rapidities
$\b_1,\ldots,\b_6$ is implied. 
Then according to the formula (\ref{F0}) 
\begin{align}
&h_0(\s_1,\s_2)=\frac{g(\s_1,\s_2)}{\s_2-\s_1-\pi i}-
\frac{g(\s_2,\s_1)}{\s_1-\s_2-\pi i}&
\label{h03}
\end{align}
The formula (\ref{fin})
contains two terms for $p=0$ and $p=1$ respectively:
\begin{align}
&\mathcal{L}(p_1,p_2)= I_0 + I_1&
\label{n=3}
\end{align}
with polynomials $p_1$ and $p_2$ which belong to the set of polynomials
$s_{-2},s_{-1},s_1,s_2$.
The term $I_0$ corresponds to the case when partitions
$$
S=\overline{ S}=\emptyset\quad T=(1,2)
$$
\begin{align}
&I_0\equiv I_{\emptyset,\emptyset,(1,2)}= 
\int_{\Gamma_1}\frac{d\s_1}{2\pi i}
\int_{\Gamma_2}\frac{d\s_2}{2\pi i}
X_1(\s_1)\Phi(\s_1) X_2(\s_2)\Phi(\s_2) h_0(\s_1,\s_2)&
\label{I03}
\end{align}
If one tries to transpose indices $1$ and $2$ in this formula:
\begin{align}
{I_0}_{\mid_{1\leftrightarrow 2}}= 
&\int_{\Gamma_1}\frac{d\s_1}{2\pi i}
\int_{\Gamma_2}\frac{d\s_2}{2\pi i}
X_2(\s_1)\Phi(\s_1) X_1(\s_2)\Phi(\s_2) h_0(\s_1,\s_2)&\non\\
&=-\int_{\Gamma_2}\frac{d\s_1}{2\pi i}
\int_{\Gamma_1}\frac{d\s_2}{2\pi i}
X_1(\s_1)\Phi(\s_1) X_2(\s_2)\Phi(\s_2) h_0(\s_1,\s_2)&\non\\
\end{align}
If it were not necessary to transpose the contours $\Gamma_1$ and
$\Gamma_2$ the last expression would be $-I_0$. Due to the transposition
of the contours we get one additional term: 
\begin{align}
&{I_0}_{\mid_{1\leftrightarrow 2}}=-I_0 + J_0&
\label{I0tr}
\end{align}
where
\begin{align}
&{J_0}=\int_{\Gamma_1}\frac{d\s_1}{2\pi i}B_{12}(\s_1)R_1(h_0(\s_1,\s_2))&
\label{J0}
\end{align}
and 
\begin{align}
&B_{12}(\s_1)=X_1(\s_1)X_2(\s_1+\pi i)+X_2(\s_1)X_1(\s_1+\pi i)&
\label{B12}
\end{align}
$$
R_1=\frac{1}{P(\s_1+\frac{\pi i}{2})}\mbox{Res}_{\s_2=\s_1+\pi i} 
$$
Here we have used the property
\begin{align}
&\Phi(\s)\Phi(\s+\pi i) = \frac{1}{P(\s+\frac{\pi i}{2})}&
\label{fifi}
\end{align}

In order to compensate this additional term $J_0$ we need
the second term $I_1$ in the formula (\ref{n=3}). In this
case there is only one partition corresponding to $p=1$ which
satisfies the conditions (\ref{SST})
$$
S=(1)\quad\overline{ S}=(2)\quad T=\emptyset
$$
\begin{align}
&I_1\equiv I_{(1),(2),\emptyset}= 
\int_{\Gamma_1}\frac{d\s_1}{2\pi i}
X_1(\s_1)p_2(\s_1)\Phi(\s_1) h_1(\s_1)&
\label{I13}
\end{align}
where
\begin{align}
&h_1(\s_1)=\d^{-1}_{\s_1}[R_1h_0(\s_1,\s_2)]&
\label{h13}
\end{align}
Using the formula (\ref{h03}) one gets
$$
[R_1h_0(\s_1,\s_2)]=\frac{g(\s_1,\s_1+\pi i)}{P(\s_1+\frac{\pi i}{2})}
$$
One can check that due to the formula (\ref{D}) the polynomial
$g(\s_1,\s_1+\pi i)$ has a divisor $P(\s_1+\frac{\pi i}{2})$ i.e.
$[R_1h_0(\s_1,\s_2)]$ is a polynomial. Therefore the operator
$\d^{-1}_{\s_1}$ is well-defined in this case and may be calculated
explicitly. 

In order to check that the term $I_1$ defined in (\ref{I13}) is suitable
for compensation of the additional term $J_0$ given by (\ref{J0})
let us substitute to the r.h.s. of (\ref{J0}) 
$$
R_1h_0(\s_1,\s_2)=\d_{\s_1}h_1(\s_1)
$$
which is an obvious consequnce of the formula (\ref{h13}). Then we get
\begin{align}
&{J_0}=\int_{\Gamma_1}\frac{d\s_1}{2\pi i}B_{12}(\s_1)\d_{\s_1}h_1(\s_1)&
\non\\
&=-\int_{\Gamma_1}\frac{d\s_1}{2\pi i}(X_1(\s_1)p_2(\s_1)+X_2(\s_1)p_1(\s_1))
\Phi(\s_1)h_1(\s_1)&
\label{J0a}
\end{align}
Actually the last expression is just
$$
-I_1 - {I_1}_{\mid_{1\leftrightarrow 2}}
$$
and we come to the result that the r.h.s. of the formula (\ref{n=3})
is the skew-symmetric function.

3. {\bf The case $n=4$}

First let us represent the function $D$ from (\ref{D}) in 
analogous to (\ref{D3}) form:
\begin{align}
&D(\s_1,\s_2,\s_3|\b_1,\ldots,\b_8)_{----++++} = &
\non\\
&
\frac{g(\s_1,\s_2,\s_3)}{(\s_2-\s_1-\pi i)(\s_3-\s_1-\pi i)(\s_3-\s_2-\pi i)}&
\label{D4}
\end{align}
where $g$ is a polynomial of both variables $\s_1,\s_2,\s_3$ and 
of the rapidities $\b_1,\ldots,\b_8$. 
An explicit form of this polynomial is defined by the formula (\ref{D}).
For us it is important that this polynomial has
the property that three polynomials \\
$g(\s_1,\s_1+\pi i,\s_3), g(\s_1,\s_3,\s_1+\pi i), g(\s_3,\s_1,\s_1+\pi i)$
contain divisor\\ 
$P(\s_1+\frac{\pi i}{2})$ while 
the polynomial $g(\s_1,\s_1+\pi i,\s_1+2\pi i)$ is divisible by\\
$P(\s_1+\frac{\pi i}{2})P(\s_1+\frac{3\pi i}{2})$.
Then according to the formula (\ref{F0}) 

\begin{align}
&h_0(\s_1,\s_2,\s_3)=&
\non\\
&
\frac{g(\s_1,\s_2,\s_3)}{(\s_{21}-\pi i)(\s_{31}-\pi i)(\s_{32}-\pi i)}-
\frac{g(\s_2,\s_1,\s_3)}{(\s_{12}-\pi i)(\s_{31}-\pi i)(\s_{32}-\pi i)}+
&
\non\\
&
\frac{g(\s_2,\s_3,\s_1)}{(\s_{32}-\pi i)(\s_{12}-\pi i)(\s_{13}-\pi i)}-
\frac{g(\s_3,\s_2,\s_1)}{(\s_{23}-\pi i)(\s_{12}-\pi i)(\s_{13}-\pi i)}+
&
\non\\
&
\frac{g(\s_3,\s_1,\s_2)}{(\s_{13}-\pi i)(\s_{23}-\pi i)(\s_{21}-\pi i)}-
\frac{g(\s_1,\s_3,\s_2)}{(\s_{31}-\pi i)(\s_{23}-\pi i)(\s_{21}-\pi i)}
&
\label{h04}
\end{align}

As in the previous case the formula (\ref{fin})
contains two terms for $p=0$ and $p=1$ respectively:
\begin{align}
&\mathcal{L}(p_1,p_2,p_3)= I_0 + I_1&
\label{n=4}
\end{align}
with polynomials $p_1, p_2, p_3$ from the set 
$s_{-3},s_{-2},s_{-1},s_1,s_2,s_3$.

The term $I_0$ corresponds to the case when partitions
$$
S=\overline{ S}=\emptyset\quad T=(1,2,3)
$$
\begin{align}
&I_0\equiv I_{\emptyset,\emptyset,(1,2,3)}= &\non\\
&\int_{\Gamma_1}\frac{d\s_1}{2\pi i}
\int_{\Gamma_2}\frac{d\s_2}{2\pi i}
\int_{\Gamma_3}\frac{d\s_3}{2\pi i}
X_1(\s_1)\Phi(\s_1) X_2(\s_2)\Phi(\s_2)X_3(\s_3)\Phi(\s_3) h_0(\s_1,\s_2,\s_3)&
\label{I04}
\end{align}

For the next term corresponding to $p=1$ there are three contributions
because one can find three different partitions that satisfy the
conditions (\ref{SST})
$$
S=(1)\quad\overline{ S}=(2)\quad T=(3)
$$
$$
S=(1)\quad\overline{ S}=(3)\quad T=(2)
$$
$$
S=(2)\quad\overline{ S}=(3)\quad T=(1)
$$
and according to the formulae (\ref{I},\ref{fin})
\begin{align}
&I_1= I_{(1),(2),(3)}-I_{(1),(3),(2)}+I_{(2),(3),(1)}&
\label{I14}
\end{align}
where
\begin{align}
&I_{(1),(2),(3)}=
\int_{\Gamma_1}\frac{d\s_1}{2\pi i}
\int_{\Gamma_3}\frac{d\s_3}{2\pi i}
X_1(\s_1)p_2(\s_1)\Phi(\s_1)X_3(\s_3)\Phi(\s_3) h_1(\s_1|\s_3)&
\non\\
&I_{(1),(3),(2)}=
\int_{\Gamma_1}\frac{d\s_1}{2\pi i}
\int_{\Gamma_2}\frac{d\s_2}{2\pi i}
X_1(\s_1)p_3(\s_1)\Phi(\s_1)X_2(\s_2)\Phi(\s_2) h_1(\s_1|\s_2)&
\non\\
&I_{(2),(3),(1)}=
\int_{\Gamma_1}\frac{d\s_1}{2\pi i}
\int_{\Gamma_2}\frac{d\s_2}{2\pi i}
X_2(\s_2)p_3(\s_2)\Phi(\s_2)X_1(\s_1)\Phi(\s_1) h_1(\s_2|\s_1)&
\label{I123}
\end{align}
and due to the formula (\ref{hp}) 
\begin{align}
&h_1(\s_1|\s_3)=
\d^{-1}_{\s_1}\biggl(R_1h_0(\s_1,\s_2,\s_3) - D_3f_1(\s_1|\s_3)\biggr)&
\label{h14}
\end{align}
with the ``exact form'' given by (\ref{D_l}) 
\begin{align}
D_3f_1(\s_1|\s_3)=
&\frac{f_1(\s_1|\s_3)}{P(\s_3+\frac{3\pi i}{2})}-
\frac{f_1(\s_1|\s_3-2\pi i)}{P(\s_3-\frac{3\pi i}{2})}&
\label{D_3}
\end{align}

In order to define the function $f_1$ we need some facts about
the structure of non-zero residues. Let us introduce the notation
$$
\mbox{Res}_{\s_2=\s_1+\pi i} h_0(\s_1,\s_2,\s_3)\equiv
V_1(\s_1|\s_3)
$$
then it is not difficult to check that
$$
\mbox{Res}_{\s_2=\s_1-\pi i} h_0(\s_1,\s_2,\s_3)=
V_1(\s_1-\pi i|\s_3)
$$
and 
$$
\mbox{Res}_{\s_2=\s_1} h_0(\s_1,\s_2,\s_3)=0
$$
Explicitly,
\begin{align}
&V_1(s_1|\s_3)=\frac{g(\s_1,\s_1+\pi i,\s_3)}
{(\s_{31}-\pi i)(\s_{31}-2\pi i)}+
\frac{g(\s_3,\s_1,\s_1+\pi i)}
{(\s_{31}-\pi i)\s_{13}}-
\frac{g(\s_1,\s_3,\s_1+\pi i)}
{(\s_{31}-\pi i)\s_{13}}&
\label{V1}
\end{align}

We shall also need
\begin{align}
\mbox{Res}_{\s_3=\s_1+2\pi i} V_1(\s_1|\s_3)\equiv
V_2(\s_1)
\label{V2p2}
\end{align}
which due to (\ref{V1}) is 
\begin{align}
&V_2(\s_1)=\frac{g(\s_1,\s_1+\pi i,\s_1+2\pi i)}{\pi i}&
\label{V2}
\end{align}
It is easy to check that
\begin{align}
\mbox{Res}_{\s_3=\s_1-\pi i} V_1(\s_1|\s_3)=
-V_2(\s_1-\pi i)
\label{V2m1}
\end{align}
and
$$
\mbox{Res}_{\s_3=\s_1+\pi i} V_1(\s_1|\s_3)=
\mbox{Res}_{\s_3=\s_1} V_1(\s_1|\s_3)=0
$$

Our claim is that 
\begin{align}
&f_1(\s_1|\s_3)=\frac{V_2(\s_3)}{\s_{13}}
\label{f_1n=4}
\end{align}

Let us briefly explain this result. First we should 
check that the r.h.s. of (\ref{h14}) is well-defined.
Indeed, using (\ref{V2p2},\ref{V2m1})
one can establish that the expression
$$
\frac{V_1(\s_1|\s_3)}{P(\s_1+\frac{\pi i}{2})}-
\frac{V_2(\s_3)}{(\s_{13}-\pi i) P(\s_3+\frac{3\pi i}{2})}+
\frac{V_2(\s_3-2\pi i)}{(\s_{13}+2\pi i)P(\s_3-\frac{3\pi i}{2})}
$$
is a polynomial of $\s_1$ i.e. the operator $\d^{-1}_{\s_1}$ is
well-defined for it. Here we also used
the above property of the polynomial $g$ which provides that
the ratio\\ 
$\frac{V_1(\s_1|\s_3)}{P(\s_1+\frac{\pi i}{2})}$
has poles only for the variable $\s_{13}$ because the numerator
$V_1(\s_1|\s_3)$ is divisible by the polynomial $P(\s_1+\frac{\pi i}{2})$.

Then if (\ref{f_1n=4}) is correct then 
the expression in brackets of the r.h.s. of (\ref{h14}) is
$$
\frac{V_1(\s_1|\s_3)}{P(\s_1+\frac{\pi i}{2})}-
\frac{V_2(\s_3)}{\s_{13}P(\s_3+\frac{3\pi i}{2})}+
\frac{V_2(\s_3-2\pi i)}{(\s_{13}+2\pi i)P(\s_3-\frac{3\pi i}{2})}
$$
which up to the above polynomial of $\s_1$ is as follows:
$$
(\frac{1}{\s_{13}-\pi i}-\frac{1}{\s_{13}})
\frac{V_2(\s_3)}{P(\s_3+\frac{3\pi i}{2})}=
-\d_{\s_1}\biggl(\frac{V_2(\s_3)}{(\s_{13}-\pi i) P(\s_3+\frac{3\pi i}{2})}
\biggr)
$$

Now let us check the skew-symmetry of the formula 
(\ref{n=4}). Actually it is enough to do for
two transpositions $1\leftrightarrow 2$ and 
$2\leftrightarrow 3$. Let us check it, for instance, for the transposition
$2\leftrightarrow 3$. 
Proceeding in analogous way as we did for the previous case
for derivation of the formula (\ref{I0tr}) we can get for the
first term $I_0$ defined by (\ref{I04})
\begin{align}
&I_0+{I_0}_{\mid_{2\leftrightarrow 3}} = J_0&
\label{I0tr4}
\end{align}
where now
\begin{align}
&{J_0}=\int_{\Gamma_1}\frac{d\s_1}{2\pi i}\int_{\Gamma_2}\frac{d\s_2}{2\pi i}
X_1(\s_1)\Phi(\s_1)B_{23}(\s_2)R_2h_0(\s_1,\s_2,\s_3)&
\label{J04}
\end{align}
and 
\begin{align}
&B_{23}(\s_2)=X_2(\s_2)X_3(\s_2+\pi i)+X_3(\s_2)X_2(\s_2+\pi i)&
\label{B23}
\end{align}
$$
R_2=\frac{1}{P(\s_2+\frac{\pi i}{2})}\mbox{Res}_{\s_3=\s_2+\pi i}
$$
We can rewrite (\ref{h14}) in the following form
\begin{align}
&
R_2h_0(\s_1,\s_2,\s_3)=\d_{\s_2}h_1(\s_2|\s_1)+D_1f_1(\s_2|\s_1)&
\label{h14a}
\end{align}
Substituting this expression into the formula (\ref{J04}) we come
to 
\begin{align}
&
J_0=J'_0+\mathcal{F}_1&
\label{J04a}
\end{align}
where
\begin{align}
&{J'_0}=\int_{\Gamma_1}\frac{d\s_1}{2\pi i}\int_{\Gamma_2}\frac{d\s_2}{2\pi i}
X_1(\s_1)\Phi(\s_1)B_{23}(\s_2)\d_{\s_2}h_1(\s_2|\s_1)&\non\\
&=-\int_{\Gamma_1}\frac{d\s_1}{2\pi i}\int_{\Gamma_2}\frac{d\s_2}{2\pi i}
X_1(\s_1)\Phi(\s_1)(X_2(\s_2)p_3(\s_2)+X_3(\s_2)p_2(\s_2))
\Phi(\s_2)h_1(\s_2|\s_1)&\non\\
&=-I_{(2),(3),(1)}-{I_{(2),(3),(1)}}_{\mid_{2\leftrightarrow 3}}&
\label{J'04}
\end{align}
\begin{align}
&\mathcal{F}_1=
\int_{\Gamma_1}\frac{d\s_1}{2\pi i}\int_{\Gamma_2}\frac{d\s_2}{2\pi i}
X_1(\s_1)\Phi(\s_1)B_{23}(\s_2)D_1f_1(\s_2|\s_1)&\non\\
&
=-\int_{\Gamma_1}\frac{d\s_1}{2\pi i}\int_{\Gamma_2}\frac{d\s_2}{2\pi i}
B_{23}(\s_2)\frac{p_1(\s_1)f_1(\s_2|\s_1)}{P(\s_1+\frac{\pi i}{2})
P(\s_1+\frac{3\pi i}{2})}
=0&\label{F1}
\end{align}
Here we have got zero because the ratio
$$
\frac{p_1(\s_1)f_1(\s_2|\s_1)}{P(\s_1+\frac{\pi i}{2})
P(\s_1+\frac{3\pi i}{2})}
$$
can have poles only for the variable $\s_{12}$ and since 
$\s_2$ belongs to the contour $\Gamma_2$ which contains
the contour $\Gamma_1$ those poles can be only outside
of the contour $\Gamma_1$.
Thus we come to the following result 
\begin{align}
&{J_0}=-I_{(2),(3),(1)}-{I_{(2),(3),(1)}}_{\mid_{2\leftrightarrow 3}}
&
\label{J04b}
\end{align}

It is left to note that the combination of 
two residual terms in the r.h.s. of 
(\ref{I14}), namely, $ I_{(1),(2),(3)}-I_{(1),(3),(2)}$
is already skew-symmetric w.r.t. the transposition
$2\leftrightarrow 3$. 

Thus we have checked the skew-symmetry for the transposition
$2\leftrightarrow 3$. For another transposition
$1\leftrightarrow 2$ it is a bit more difficult but also possible to do.
The strategy is more or less repeats that for the
transposition $2\leftrightarrow 3$. We shall not do it here.

4. {\bf The case $n=5$}

It is straightforward to write down the function $D$ from (\ref{D}) in 
the form like (\ref{D3}) for $n=3$ and (\ref{D4}) for $n=4$
defining a polynomial $g(\s_1,\s_2,\s_3,\s_4)$ which also depends
on the rapidities $\b_1,\ldots,\b_{10}$. 
Again the main property of this polynomial is that all polynomails
of the form
$g(\ldots,\s_1,\ldots,\s_1+\pi i,\ldots)$
contain as a divisor\\ 
$P(\s_1+\frac{\pi i}{2})$,  
the polynomials $g(\ldots,\s_1,\ldots,
\s_1+\pi i,\ldots,\s_1+2\pi i)$ are divisible by
$P(\s_1+\frac{\pi i}{2})P(\s_1+\frac{3\pi i}{2})$
etc.

Now the formula (\ref{fin}) will
contain three terms for $p=0, p=1$ and $p=2$:
\begin{align}
&\mathcal{L}(p_1,p_2,p_3,p_4)= I_0 + I_1 +I_2&
\label{n=5}
\end{align}
with polynomials $p_1, p_2, p_3,p_4$ from the set 
$s_{-4},s_{-3},s_{-2},s_{-1},s_1,s_2,s_3,s_4$.

The main term $I_0$ corresponds to the partitions
$$
S=\overline{ S}=\emptyset\quad T=(1,2,3,4)
$$
\begin{align}
&I_0\equiv I_{\emptyset,\emptyset,(1,2,3,4)}= &\non\\
&\int_{\Gamma_1}\frac{d\s_1}{2\pi i}
\int_{\Gamma_2}\frac{d\s_2}{2\pi i}
\int_{\Gamma_3}\frac{d\s_3}{2\pi i}
\int_{\Gamma_4}\frac{d\s_4}{2\pi i}
X_1(\s_1)\Phi(\s_1) X_2(\s_2)\Phi(\s_2)&
\non\\
&\cdot
X_3(\s_3)\Phi(\s_3) 
X_4(\s_4)\Phi(\s_4)h_0(\s_1,\s_2,\s_3,\s_4)&
\label{I05}
\end{align}
where as above the function 
$h_0(\s_1,\ldots,\s_4)$ is defined by the formula (\ref{F0}). 

For the next term with $p=1$ there are six contributions
corresponding to six different partitions that satisfy the
conditions (\ref{SST})
$$
S=(1)\quad\overline{ S}=(2)\quad T=(34)
$$
$$
S=(1)\quad\overline{ S}=(3)\quad T=(24)
$$
$$
S=(1)\quad\overline{ S}=(4)\quad T=(23)
$$
$$
S=(2)\quad\overline{ S}=(3)\quad T=(14)
$$
$$
S=(2)\quad\overline{ S}=(4)\quad T=(13)
$$
$$
S=(3)\quad\overline{ S}=(4)\quad T=(12)
$$
and according to the formulae (\ref{I},\ref{fin})
\begin{align}
&I_1= I_{(1),(2),(34)}-I_{(1),(3),(24)}+I_{(1),(4),(23)}
+I_{(2),(3),(14)}-I_{(2),(4),(13)}+I_{(3),(4),(12)}
&
\label{I15}
\end{align}
where
\begin{align}
&I_{(i_1),(i_2),(i_3\,i_4)}=
\int_{\Gamma_{i_1}}\frac{d\s_{i_1}}{2\pi i}
\int_{\Gamma_{i_3}}\frac{d\s_{i_3}}{2\pi i}
\int_{\Gamma_{i_4}}\frac{d\s_{i_4}}{2\pi i}
X_{i_1}(\s_{i_1})p_{i_2}(\s_{i_1})\Phi(\s_{i_1})&
\non\\
&\cdot
X_{i_3}(\s_{i_3})\Phi(\s_{i_3})X_{i_4}(\s_{i_4})\Phi(\s_{i_4}) 
h_1(\s_{i_1}|\s_{i_3},\s_{i_4})&
\label{I1234}
\end{align}
and due to the formula (\ref{hp}) 
\begin{align}
&h_1(\s_1|\s_3,\s_4)=
\d^{-1}_{\s_1}\biggl(R_1h_0(\s_1,\s_2,\s_3,\s_4) - 
D_3f_1(\s_1|\s_3|\s_4)+D_4f_1(\s_1|\s_4|\s_3)\biggr)&
\label{h15}
\end{align}
with the ``exact forms'' $D_3$ and $D_4$ defined by (\ref{D_l}).
Let us note that the function $h_1(\s_1|\s_3,\s_4)$ is
manifestly anti-symmetric w.r.t. the transposition 
$\s_3\leftrightarrow\s_4$.

For the last term with $p=2$ there are three contributions
which correspond to the three possible partitions that satisfy the
conditions (\ref{SST})
$$
S=(12)\quad\overline{ S}=(34)\quad T=\emptyset
$$
$$
S=(12)\quad\overline{ S}=(43)\quad T=\emptyset
$$
$$
S=(13)\quad\overline{ S}=(24)\quad T=\emptyset
$$
and
\begin{align}
&I_2= -I_{(12),(34),\emptyset}+I_{(12),(43),\emptyset}
+I_{(13),(24),\emptyset}
&
\label{I25}
\end{align}
where
\begin{align}
&I_{(i_1\,i_2),(i_3\,i_4),\emptyset}=
\int_{\Gamma_{i_1}}\frac{d\s_{i_1}}{2\pi i}
\int_{\Gamma_{i_2}}\frac{d\s_{i_2}}{2\pi i}
X_{i_1}(\s_{i_1})p_{i_3}(\s_{i_1})\Phi(\s_{i_1})&\non\\
&
\cdot X_{i_2}(\s_{i_2})
p_{i_4}(\s_{i_2})\Phi(\s_{i_2})
h_2(\s_{i_1},\s_{i_2})&
\label{I21234}
\end{align}
and due to the formula (\ref{hp}) 
\begin{align}
&h_2(\s_1,\s_3)=
\d^{-1}_{\s_1}\d^{-1}_{\s_3}\biggl(R_1 R_3h_0(\s_1,\s_2,\s_3,\s_4) &\non\\
&-R_1[D_1f_1(\s_3|\s_1|\s_2)-D_2f_1(\s_3|\s_2|\s_1)]&\non\\
&- R_3[D_3f_1(\s_1|\s_3|\s_4)-D_4f_1(\s_1|\s_4|\s_3)]
\biggr)&
\label{h25}
\end{align}
Actually this formula for $h_2(\s_1,\s_3)$ is symmetric
w.r.t. the transposition $\s_1\leftrightarrow\s_3$.

In order to define the function $f_1$ we need notation which are
similar to those for the case $n=4$. Let us write only non-zero
residues
$$
\mbox{Res}_{\s_2=\s_1+\pi i} h_0(\s_1,\s_2,\s_3,\s_4)\equiv
V_1(\s_1|\s_3,\s_4)
$$
$$
\mbox{Res}_{\s_2=\s_1-\pi i} h_0(\s_1,\s_2,\s_3,\s_4)=
V_1(\s_1-\pi i|\s_3,\s_4)
$$
\begin{align}
&\mbox{Res}_{\s_3=\s_1+2\pi i} V_1(\s_1|\s_3,\s_4)\equiv
V_2(\s_1|\s_4)&\non\\
&\mbox{Res}_{\s_3=\s_1-\pi i} V_1(\s_1|\s_3,\s_4)=
-V_2(\s_1-\pi i|\s_4)&
\label{V2n=5}
\end{align}
\begin{align}
&\mbox{Res}_{\s_4=\s_1+3\pi i} V_2(\s_1|\s_4)\equiv
V_3(\s_1)&\non\\
&\mbox{Res}_{\s_4=\s_1+\pi i} V_2(\s_1|\s_4)\equiv
V_{2,1}(\s_1)&\non\\
&\mbox{Res}_{\s_4=\s_1-\pi i} V_2(\s_1|\s_4)=
V_3(\s_1-\pi i)&
\label{V3n=5}
\end{align}

The result for the function $f_1$ looks as follows:
\begin{align}
f_1(\s_1|\s_3|\s_4)=
\frac{V_2(\s_3|\s_4)}{\s_{13}}+(\frac{1}{\s_{13}}-
\frac{1}{\s_{13}-\pi i})\frac{V_3(\s_3)}{\s_{34}+3\pi i}
\label{f_1n=5}
\end{align}

An important property of the function $f_1$ is that
non-zero residues w.r.t. to the pair of the
variables $\s_3,\s_4$ are only in three points 
$\s_4=\s_3+3\pi i,\s_4=\s_3+\pi i,\s_4=\s_3-\pi i$.
Let us denote
\begin{align}
&\mbox{Res}_{\s_4=\s_3+3\pi i} f_1(\s_1|\s_3|\s_4)\equiv
\chi_3(\s_1|\s_3)&\non\\
&\mbox{Res}_{\s_4=\s_3+\pi i} f_1(\s_1|\s_3|\s_4)\equiv
\chi_1(\s_1|\s_3)&
\label{f_1prop}
\end{align}
One can also check that
\begin{align}
&\mbox{Res}_{\s_4=\s_3-\pi i} f_1(\s_1|\s_3|\s_4)=
\chi_3(\s_1|\s_3-\pi i)&
\label{f_1prop1}
\end{align}

First we should be convinced that the expressions (\ref{h15}) and
(\ref{h25}) for the functions $h_1$ and $h_2$ respectively
are well-defined. For the case of the function $h_1$ one can
proceed in the same line as it was done above for $n=4$.
In order to check that (\ref{h25}) is well-defined also we
need the following relation which can be verified straightforwardly
using the above properties (\ref{f_1prop},\ref{f_1prop1}) of
the function $f_1$
\begin{align}
&R_1\biggl(D_1f_1(\s_3|\s_1|\s_2)-D_2f_1(\s_3|\s_2|\s_1)\biggr)
=\d_{\s_1}\biggl(
\frac{\chi_1(\s_3|\s_1-\pi i)}{P(\s_1-\frac{\pi i}{2})P(\s_1+\frac{\pi i}{2})}
&\non\\
&
+
\frac{\chi_3(\s_3|\s_1-\pi i)}{P(\s_1-\frac{\pi i}{2})P(\s_1+\frac{3\pi i}{2})}
+
\frac{\chi_3(\s_3|\s_1-2\pi i)}
{P(\s_1-\frac{3\pi i}{2})P(\s_1+\frac{\pi i}{2})}\biggr)&
\label{f_1prop2}
\end{align}

Then the bracket in the r.h.s. of (\ref{h25}) can be rewritten
as follows
\begin{align}
&R_3\biggl(R_1h_0(\s_1,\s_2,\s_3,\s_4)-
D_3f_1(\s_1|\s_3|\s_4)+D_4f_1(\s_1|\s_4|\s_3)\biggr)
&\non\\
&-R_1\biggl(D_1f_1(\s_3|\s_1|\s_2)-D_2f_1(\s_3|\s_2|\s_1)\biggr)&
\label{h25brac}
\end{align}
For the first term the operator $\d^{-1}_{\s_1}$ is well-defined
just because 
it is $\d_{\s_1}R_3h_1(\s_1|\s_3,\s_4)$ 
according to the formula (\ref{h15}). For the second term
in the formula 
(\ref{h25brac}) $\d^{-1}_{\s_1}$ does exist due to the
identity (\ref{f_1prop2}). Thus we have proved that the
operator $\d^{-1}_{\s_1}$ is well-defined for the whole expression
(\ref{h25brac}). On the other hand it is manifestly symmetric
w.r.t. the transposition $\s_1\leftrightarrow\s_3$. It means
that the operator $\d^{-1}_{\s_3}$ is well-defined as well.

Let us check the skew-symmetry of (\ref{n=5}) for the transposition
$3\leftrightarrow 4$. 
An analog of the formulae (\ref{I0tr}) for $n=3$ and (\ref{I0tr4}) for
$n=4$ for the first term $I_0$ defined by (\ref{I05}) is
\begin{align}
&I_0+{I_0}_{\mid_{3\leftrightarrow 4}} = J_0&
\label{I0tr5}
\end{align}
with a new
\begin{align}
&{J_0}=
\int_{\Gamma_1}\frac{d\s_1}{2\pi i}\int_{\Gamma_2}\frac{d\s_2}{2\pi i}
\int_{\Gamma_3}\frac{d\s_3}{2\pi i}
X_1(\s_1)\Phi(\s_1)X_2(\s_2)\Phi(\s_2)
B_{34}(\s_3)&\non\\
&\cdot R_3h_0(\s_1,\s_2,\s_3,\s_4)&
\label{J05}
\end{align}
and analogously to (\ref{B12}) and (\ref{B23})
\begin{align}
&B_{34}(\s_2)=X_3(\s_3)X_4(\s_3+\pi i)+X_4(\s_3)X_3(\s_3+\pi i)&
\label{B34}
\end{align}
$$
R_3=\frac{1}{P(\s_3+\frac{\pi i}{2})}\mbox{Res}_{\s_4=\s_3+\pi i}
$$
As we did above we rewrite (\ref{h15}) in the following form
\begin{align}
&
R_3h_0(\s_1,\s_2,\s_3,\s_4)=\d_{\s_3}h_1(\s_3|\s_1,\s_2)+
D_1f_1(\s_3|\s_1|\s_2)-D_2f_1(\s_3|\s_2|\s_1)&
\label{h15a}
\end{align}
Substituting this expression into the formula (\ref{J05}) we get
\begin{align}
&
J_0=J'_0+\mathcal{F}_1+\mathcal{F}_2&
\label{J05a}
\end{align}
where
\begin{align}
&{J'_0}=
\int_{\Gamma_1}\frac{d\s_1}{2\pi i}\int_{\Gamma_2}\frac{d\s_2}{2\pi i}
\int_{\Gamma_3}\frac{d\s_3}{2\pi i}
X_1(\s_1)\Phi(\s_1)X_2(\s_2)\Phi(\s_2)
B_{34}(\s_3)\d_{\s_3}h_1(\s_3|\s_1,\s_2)&\non\\
&=-\int_{\Gamma_1}\frac{d\s_1}{2\pi i}\int_{\Gamma_2}\frac{d\s_2}{2\pi i}
\int_{\Gamma_3}\frac{d\s_3}{2\pi i}
X_1(\s_1)\Phi(\s_1)X_2(\s_2)\Phi(\s_2)&\non\\
&\cdot (X_3(\s_3)p_4(\s_3)+X_4(\s_3)p_3(\s_3))
\Phi(\s_3)h_1(\s_3|\s_1,\s_2)&\non\\
&=-I_{(3),(4),(12)}-{I_{(3),(4),(12)}}_{\mid_{3\leftrightarrow 4}}&
\label{J'05}
\end{align}
\begin{align}
&\mathcal{F}_1=
\int_{\Gamma_1}\frac{d\s_1}{2\pi i}\int_{\Gamma_2}\frac{d\s_2}{2\pi i}
\int_{\Gamma_3}\frac{d\s_3}{2\pi i}
X_1(\s_1)\Phi(\s_1)X_2(\s_2)\Phi(\s_2)B_{34}(\s_3)
D_1f_1(\s_3|\s_1|\s_2)&\non\\
&
=-\int_{\Gamma_1}\frac{d\s_1}{2\pi i}\int_{\Gamma_2}\frac{d\s_2}{2\pi i}
\int_{\Gamma_3}\frac{d\s_3}{2\pi i}X_2(\s_2)\Phi(\s_2)
B_{34}(\s_3)\frac{p_1(\s_1)f_1(\s_3|\s_1|\s_2)}{P(\s_1+\frac{\pi i}{2})
P(\s_1+\frac{3\pi i}{2})}
=0&\label{F1n=5}
\end{align}
The reason why we get zero here is similar to that of the
previous case, namely, because the ratio
$$
\frac{p_1(\s_1)f_1(\s_3|\s_1|\s_2)}{P(\s_1+\frac{\pi i}{2})
P(\s_1+\frac{3\pi i}{2})}
$$
can have poles only for the variables $\s_{12}$ and $\s_{13}$.
In both cases these poles are outside of the contour $\Gamma_1$.

The situation with the third term in (\ref{J05a}) $\mathcal{F}_2$
is different
\begin{align}
&\mathcal{F}_2=
-\int_{\Gamma_1}\frac{d\s_1}{2\pi i}\int_{\Gamma_2}\frac{d\s_2}{2\pi i}
\int_{\Gamma_3}\frac{d\s_3}{2\pi i}
X_1(\s_1)\Phi(\s_1)X_2(\s_2)\Phi(\s_2)B_{34}(\s_3)
D_2f_1(\s_3|\s_2|\s_1)&\non\\
&
=\int_{\Gamma_1}\frac{d\s_1}{2\pi i}\int_{\Gamma_2}\frac{d\s_2}{2\pi i}
\int_{\Gamma_3}\frac{d\s_3}{2\pi i}X_1(\s_1)\Phi(\s_1)
B_{34}(\s_3)\frac{p_2(\s_2)f_1(\s_3|\s_2|\s_1)}{P(\s_2+\frac{\pi i}{2})
P(\s_2+\frac{3\pi i}{2})}
&\label{F2n=5}
\end{align}
This is not zero because now there is a contribution from the poles
for the variable $\s_{12}$. Using the properties of the function 
$f_1$ (\ref{f_1prop},\ref{f_1prop1}) we can get
\begin{align}
&
\int_{\Gamma_2}\frac{d\s_2}{2\pi i}
\frac{p_2(\s_2)f_1(\s_3|\s_2|\s_1)}{P(\s_2+\frac{\pi i}{2})
P(\s_2+\frac{3\pi i}{2})}&\non\\
&
=D_1\biggl(p_2(\s_1)
\frac{\chi_3(\s_3|\s_1-\pi i)}{P(\s_1-\frac{\pi i}{2})}-
p_2(\s_1+2\pi i)\frac{\chi_3(\s_3|\s_1)}{P(\s_1+\frac{5\pi i}{2})}\biggr)
&\non\\
&
-p_2(\s_1)
\biggl(
\frac{\chi_1(\s_3|\s_1-\pi i)}{P(\s_1-\frac{\pi i}{2})P(\s_1+\frac{\pi i}{2})}
+
\frac{\chi_3(\s_3|\s_1-\pi i)}{P(\s_1-\frac{\pi i}{2})P(\s_1+\frac{3\pi i}{2})}
&\non\\
&
+
\frac{\chi_3(\s_3|\s_1-2\pi i)}
{P(\s_1-\frac{3\pi i}{2})P(\s_1+\frac{\pi i}{2})}\biggr)
&\non\\
&
=D_1\biggl(p_2(\s_1)
\frac{\chi_3(\s_3|\s_1-\pi i)}{P(\s_1-\frac{\pi i}{2})}-
p_2(\s_1+2\pi i)\frac{\chi_3(\s_3|\s_1)}{P(\s_1+\frac{5\pi i}{2})}\biggr)
&\non\\
&
-p_2(\s_1)
\d^{-1}_{\s_1}R_1\biggl(D_1f_1(\s_3|\s_1|\s_2)-D_2f_1(\s_3|\s_2|\s_1)\biggr)
&
\label{D2res}
\end{align}
The last equality follows from the relation (\ref{f_1prop2}).
Substituting this result for the integration over $\s_2$ into
(\ref{F2n=5}) and using the fact that the term $D_1(\ldots)$
in (\ref{D2res}) 
does not contribute one arrives at
\begin{align}
&\mathcal{F}_2=
\int_{\Gamma_1}\frac{d\s_1}{2\pi i}
\int_{\Gamma_3}\frac{d\s_3}{2\pi i}
X_1(\s_1)p_2(\s_1)\Phi(\s_1)B_{34}(\s_3)
&\non\\
&
\cdot
\d^{-1}_{\s_1}R_1\biggl(-D_1f_1(\s_3|\s_1|\s_2)+D_2f_1(\s_3|\s_2|\s_1)\biggr)
&
\label{F2n=5a}
\end{align}

Thus we come to the following result 
\begin{align}
&{J_0}=-I_{(3),(4),(12)}-{I_{(3),(4),(12)}}_{\mid_{3\leftrightarrow 4}}
+\mathcal{F}_2
&
\label{J05b}
\end{align}
with $\mathcal{F}_2$ given by (\ref{F2n=5a}).

Now let us note that the part of (\ref{I15})
$$
-I_{(1),(3),(24)}+I_{(1),(4),(23)}
+I_{(2),(3),(14)}-I_{(2),(4),(13)}
$$
is already skew-symmetric w.r.t. the transposition 
$3\leftrightarrow 4$. It is left to treat the
very first term in (\ref{I15})
$$
I_{(1),(2),(34)}
$$
Proceeding in a similar way as we treated above the term $I_0$
by the derivation of the formula (\ref{J05}) and using
the formula (\ref{h15}) for the function $h_1$ we can obtain
\begin{align}
&I_{(1),(2),(34)}+{I_{(1),(2),(34)}}_{\mid_{3\leftrightarrow 4}} +
\mathcal{F}_2 = J_1&
\label{I1234tr5}
\end{align}
where
\begin{align}
&{J_1}=
\int_{\Gamma_1}\frac{d\s_1}{2\pi i}
\int_{\Gamma_3}\frac{d\s_3}{2\pi i}
X_1(\s_1)p_2(\s_1)\Phi(\s_1)
B_{34}(\s_3)\d^{-1}_{\s_1}\biggl(R_1 R_3h_0(\s_1,\s_2,\s_3,\s_4)
&\non\\ 
&
+R_1[-D_1f_1(\s_3|\s_1|\s_2)+D_2f_1(\s_3|\s_2|\s_1)]
&\non\\
&
+R_3[-D_3f_1(\s_1|\s_3|\s_4)+D_4f_1(\s_1|\s_4|\s_3)]
\biggr)&
\label{J15}
\end{align}
Actually, expression in the curled bracket coincides with
that from the formula (\ref{h25}). Therefore
\begin{align}
&{J_1}=
\int_{\Gamma_1}\frac{d\s_1}{2\pi i}
\int_{\Gamma_3}\frac{d\s_3}{2\pi i}
X_1(\s_1)p_2(\s_1)\Phi(\s_1)
B_{34}(\s_3)\d_{\s_3}h_2(\s_1,\s_3)
&\non\\
&
=-\int_{\Gamma_1}\frac{d\s_1}{2\pi i}
\int_{\Gamma_3}\frac{d\s_3}{2\pi i}
X_1(\s_1)p_2(\s_1)\Phi(\s_1)
&\non\\
&
\cdot (X_3(\s_3)p_4(\s_3)+X_4(\s_3)p_3(\s_3))\Phi(\s_3)h_2(\s_1,\s_3)
&\non\\
&=-I_{(13),(24),\emptyset}-
{I_{(13),(24),\emptyset}}_{\mid_{3\leftrightarrow 4}}
&
\label{J15a}
\end{align}

What is left is to note that the part of (\ref{I25}) which contains
the residual two terms
$$
-I_{(12),(34),\emptyset}+I_{(12),(43),\emptyset}
$$
is already skew-symmetric w.r.t. the transposition 
$3\leftrightarrow 4$.

Thus we have checked the skew-symmetry of the whole expression
(\ref{n=5}) for the transposition
$3\leftrightarrow 4$. For other transpositions
$1\leftrightarrow 2$ and $2\leftrightarrow 3$
this can also be done.

5. {\bf The case $n=6$}

As in the previous case
the formula (\ref{fin}) 
contains three terms for $p=0, p=1$ and $p=2$:
\begin{align}
&\mathcal{L}(p_1,p_2,p_3,p_4,p_5)= I_0 + I_1 +I_2&
\label{n=6}
\end{align}
where $p_1, p_2, p_3,p_4,p_5$ belong to the set 
of polynomials\\
$s_{-5},\ldots,s_{-2},s_{-1},s_1,s_2,\ldots,s_5$.

The main term $I_0$ corresponds to the partitions
$$
S=\overline{ S}=\emptyset\quad T=(1,2,3,4,5)
$$
\begin{align}
&I_0\equiv I_{\emptyset,\emptyset,(1,2,3,4,5)}= &\non\\
&\int_{\Gamma_1}\frac{d\s_1}{2\pi i}
\ldots
\int_{\Gamma_5}\frac{d\s_5}{2\pi i}
X_1(\s_1)\Phi(\s_1) X_2(\s_2)\Phi(\s_2)X_3(\s_3)\Phi(\s_3)&
\non\\
&
\cdot X_4(\s_4)\Phi(\s_4) 
X_5(\s_5)\Phi(\s_5)h_0(\s_1,\s_2,\s_3,\s_4,\s_5)&
\label{I06}
\end{align}
where as above the function 
$h_0(\s_1,\ldots,\s_5)$ is defined by the formula (\ref{F0}). 

For the next term with $p=1$ there are ten contributions
corresponding to ten partitions 
$$
S=(1)\quad\overline{ S}=(2)\quad T=(345)
$$
$$
S=(1)\quad\overline{ S}=(3)\quad T=(245)
$$
$$
S=(1)\quad\overline{ S}=(4)\quad T=(235)
$$
$$
S=(1)\quad\overline{ S}=(5)\quad T=(234)
$$
$$
S=(2)\quad\overline{ S}=(3)\quad T=(145)
$$
$$
S=(2)\quad\overline{ S}=(4)\quad T=(135)
$$
$$
S=(2)\quad\overline{ S}=(5)\quad T=(134)
$$
$$
S=(3)\quad\overline{ S}=(4)\quad T=(125)
$$
$$
S=(3)\quad\overline{ S}=(5)\quad T=(124)
$$
$$
S=(4)\quad\overline{ S}=(5)\quad T=(123)
$$
and according to the formulae (\ref{I},\ref{fin})
\begin{align}
&I_1= I_{(1),(2),(345)}-I_{(1),(3),(245)}+I_{(1),(4),(235)}-
I_{(1),(5),(234)}+
I_{(2),(3),(145)}&\non\\
&-I_{(2),(4),(135)}+I_{(2),(5),(134)}+
I_{(3),(4),(125)}-I_{(3),(5),(124)}+I_{(4),(5),(123)}
&
\label{I16}
\end{align}
where
\begin{align}
&I_{(i_1),(i_2),(i_3\,i_4\,i_5)}=
\int_{\Gamma_{i_1}}\frac{d\s_{i_1}}{2\pi i}
\int_{\Gamma_{i_3}}\frac{d\s_{i_3}}{2\pi i}
\int_{\Gamma_{i_4}}\frac{d\s_{i_4}}{2\pi i}
\int_{\Gamma_{i_5}}\frac{d\s_{i_5}}{2\pi i}
X_{i_1}(\s_{i_1})p_{i_2}(\s_{i_1})\Phi(\s_{i_1})&\non\\
&
\cdot X_{i_3}(\s_{i_3})\Phi(\s_{i_3})
X_{i_4}(\s_{i_4})\Phi(\s_{i_4})
X_{i_5}(\s_{i_5})\Phi(\s_{i_5})h_1(\s_{i_1}|\s_{i_3},\s_{i_4},\s_{i_5})
\label{I12345}
\end{align}
and due to the formula (\ref{hp}) 
\begin{align}
&h_1(\s_1|\s_3,\s_4,\s_5)=
\d^{-1}_{\s_1}\biggl(R_1h_0(\s_1,\s_2,\s_3,\s_4,\s_5) &\non\\
&- 
D_3f_1(\s_1|\s_3|\s_4,\s_5)+D_4f_1(\s_1|\s_4|\s_3,\s_5)-
D_5f_1(\s_1|\s_5|\s_3,\s_4)\biggr)&
\label{h16}
\end{align}
Obviously the function $h_1(\s_1|\s_3,\s_4,\s_5)$ is
fully anti-symmetric for the variables $\s_3,\s_4,\s_5$.

For the last term with $p=2$ there are fifteen contributions
corresponding to all  possible partitions that satisfy the
conditions (\ref{SST})
$$
S=(12)\quad\overline{ S}=(34)\quad T=(5)
$$
$$
S=(12)\quad\overline{ S}=(43)\quad T=(5)
$$
$$
S=(12)\quad\overline{ S}=(35)\quad T=(4)
$$
$$
S=(12)\quad\overline{ S}=(53)\quad T=(4)
$$
$$
S=(12)\quad\overline{ S}=(45)\quad T=(3)
$$
$$
S=(12)\quad\overline{ S}=(54)\quad T=(3)
$$
$$
S=(13)\quad\overline{ S}=(24)\quad T=(5)
$$
$$
S=(13)\quad\overline{ S}=(25)\quad T=(4)
$$
$$
S=(13)\quad\overline{ S}=(45)\quad T=(2)
$$
$$
S=(13)\quad\overline{ S}=(54)\quad T=(2)
$$
$$
S=(14)\quad\overline{ S}=(25)\quad T=(3)
$$
$$
S=(14)\quad\overline{ S}=(35)\quad T=(2)
$$
$$
S=(23)\quad\overline{ S}=(45)\quad T=(1)
$$
$$
S=(23)\quad\overline{ S}=(54)\quad T=(1)
$$
$$
S=(24)\quad\overline{ S}=(35)\quad T=(1)
$$
and
\begin{align}
&I_2= -I_{(12),(34),(5)}+I_{(12),(43),(5)}+I_{(12),(35),(4)}-
I_{(12),(53),(4)} -I_{(12),(45),(3)}
&\non\\
&
+I_{(12),(54),(3)}+I_{(13),(24),(5)}
-I_{(13),(25),(4)}+I_{(13),(45),(2)}-I_{(13),(54),(2)}
&\non\\
&
+
I_{(14),(25),(3)}-I_{(14),(35),(2)}
-I_{(23),(45),(1)}+I_{(23),(54),(1)}+I_{(24),(35),(1)}
&
\label{I26}
\end{align}
where
\begin{align}
&I_{(i_1\,i_2),(i_3\,i_4),(i_5)}=
\int_{\Gamma_{i_1}}\frac{d\s_{i_1}}{2\pi i}
\int_{\Gamma_{i_2}}\frac{d\s_{i_2}}{2\pi i}
\int_{\Gamma_{i_5}}\frac{d\s_{i_5}}{2\pi i}
X_{i_1}(\s_{i_1})p_{i_3}(\s_{i_1})\Phi(\s_{i_1})&\non\\
&
X_{i_2}(\s_{i_2})p_{i_4}(\s_{i_2})\Phi(\s_{i_2})
X_{i_5}(\s_{i_5})\Phi(\s_{i_5})
h_2(\s_{i_1},\s_{i_2}|\s_{i_5})&
\label{I212345}
\end{align}
and due to the formula (\ref{hp}) 
\begin{align}
&h_2(\s_1,\s_3|\s_5)=
\d^{-1}_{\s_1}\d^{-1}_{\s_3}\biggl(R_1 R_3h_0(\s_1,\s_2,\s_3,\s_4,\s_5) &\non\\
&-R_1[D_1f_1(\s_3|\s_1|\s_2,\s_5)-D_2f_1(\s_3|\s_2|\s_1,\s_5)+
D_5f_1(\s_3|\s_5|\s_1,\s_2)]&\non\\
&- R_3[D_3f_1(\s_1|\s_3|\s_4,\s_5)-D_4f_1(\s_1|\s_4|\s_3,\s_5)+
D_5f_1(\s_1|\s_5|\s_3,\s_4)]
&\non\\
&+D_5f_2(\s_1,\s_3|\s_5)
\biggr)&
\label{h26}
\end{align}
Like in the case $n=5$ the formula for $h_2(\s_1,\s_3|\s_5)$ is symmetric
w.r.t. the transposition $\s_1\leftrightarrow\s_3$.

In order to define the functions $f_1$ and $f_2$ we need some residues
which, in principle, may be computed explicitly using the formula (\ref{F0})
$$
\mbox{Res}_{\s_2=\s_1+\pi i} h_0(\s_1,\s_2,\s_3,\s_4,\s_5)=
V_1(\s_1|\s_3,\s_4,\s_5)
$$
$$
\mbox{Res}_{\s_2=\s_1-\pi i} h_0(\s_1,\s_2,\s_3,\s_4,\s_5)=
V_1(\s_1-\pi i|\s_3,\s_4,\s_5)
$$

$$
\mbox{Res}_{\s_3=\s_1+2\pi i} V_1(\s_1|\s_3,\s_4,\s_5)=
V_2(\s_1|\s_4,\s_5)
$$
$$
\mbox{Res}_{\s_3=\s_1-\pi i} V_1(\s_1|\s_3,\s_4,\s_5)=
-V_2(\s_1-\pi i|\s_4,\s_5)
$$

$$
\mbox{Res}_{\s_4=\s_1+3\pi i} V_2(\s_1|\s_4,\s_5)=
V_3(\s_1|\s_5)
$$
$$
\mbox{Res}_{\s_4=\s_1+\pi i} V_2(\s_1|\s_4,\s_5)=
V_{2,1}(\s_1|\s_5)
$$
$$
\mbox{Res}_{\s_4=\s_1-\pi i} V_2(\s_1|\s_4,\s_5)=
V_3(\s_1-\pi i|\s_5)
$$
$$
\mbox{Res}_{\s_5=\s_4+\pi i} V_2(\s_1|\s_4,\s_5)=
{\tilde V}_2(\s_1|\s_4)
$$

$$
\mbox{Res}_{\s_5=\s_1+4\pi i} V_3(\s_1|\s_5)=
V_4(\s_1)
$$
$$
\mbox{Res}_{\s_5=\s_1+2\pi i} V_3(\s_1|\s_5)=
V_{3,2}(\s_1)
$$
$$
\mbox{Res}_{\s_5=\s_1+\pi i} V_3(\s_1|\s_5)=
V_{3,1}(\s_1)
$$
$$
\mbox{Res}_{\s_5=\s_1-\pi i} V_3(\s_1|\s_5)=
-V_4(\s_1-\pi i)
$$

$$
\mbox{Res}_{\s_4=\s_3+3\pi i} {\tilde V}_2(\s_3|\s_4)=
V_4(\s_3)
$$
$$
\mbox{Res}_{\s_4=\s_3+2\pi i} {\tilde V}_2(\s_3|\s_4)=
V_{3,2}(\s_3)
$$
$$
\mbox{Res}_{\s_4=\s_3+\pi i} {\tilde V}_2(\s_3|\s_4)=
{\tilde V}_{2,1}(\s_3)
$$
$$
\mbox{Res}_{\s_4=\s_3} {\tilde V}_2(\s_3|\s_4)=
{\tilde V}_{2,0}(\s_3)
$$
$$
\mbox{Res}_{\s_4=\s_3-\pi i} {\tilde V}_2(\s_3|\s_4)=
V_{3,1}(\s_3-\pi i)
$$
$$
\mbox{Res}_{\s_4=\s_3-2\pi i} {\tilde V}_2(\s_3|\s_4)=
-V_4(\s_3-2\pi i)
$$

The result for the function $f_1$ looks as follows:
\begin{align}
&f_1(\s_1|\s_3|\s_4,\s_5)=\frac{V_2(\s_1|\s_4,\s_5)}{\s_{13}}&\non\\
&
-(\frac{1}{\s_{13}}-\frac{1}{\s_{13}-\pi i})p(\s_3|\s_4,\s_5)-
(\frac{1}{\s_{13}-\pi i}-\frac{1}{\s_{13}-2\pi i})q(\s_3|\s_4,\s_5)
&
\label{f1n=6}
\end{align}
where
\begin{align}
&
p(\s_3|\s_4,\s_5)=
\frac{V_3(\s_3|\s_4)}{\s_{35}+3\pi i}-
\frac{V_3(\s_3|\s_5)}{\s_{34}+3\pi i}-
&\non\\
&
-\frac{V_4(\s_3)}{(\s_{35}+4\pi i)(\s_{54}-\pi i)}+
\frac{V_4(\s_3)}{(\s_{34}+4\pi i)(\s_{45}-\pi i)}+
&\non\\
&
+\frac{V_{3,2}(\s_3)}{(\s_{35}+2\pi i)(\s_{45}-\pi i)}
-\frac{V_{3,2}(\s_3)}{(\s_{34}+2\pi i)(\s_{54}-\pi i)}
\label{p}
\end{align}
and
\begin{align}
&
q(\s_3|\s_4,\s_5)=
\frac{V_4(\s_3)}{(\s_{35}+3\pi i)(\s_{45}-\pi i)}-
\frac{V_4(\s_3)}{(\s_{34}+3\pi i)(\s_{54}-\pi i)}-
&\non\\
&
-\frac{V_4(\s_3)}{(\s_{35}+\pi i)(\s_{45}-\pi i)}
\frac{P(\s_3+\frac{3\pi i}{2})}{P(\s_3+\frac{7\pi i}{2})}+
\frac{V_4(\s_3)}{(\s_{34}+\pi i)(\s_{54}-\pi i)}
\frac{P(\s_3+\frac{3\pi i}{2})}{P(\s_3+\frac{7\pi i}{2})}
\label{q}
\end{align}

The answer for the function $f_2$ is
\begin{align}
&f_2(\s_1,\s_3|\s_5)=
\frac{V_4(\s_5)}{(\s_{15}-\pi i)(\s_{35}-\pi i)P(\s_5+\frac{7\pi i}{2})}-
\frac{V_4(\s_5)}{\s_{15}\s_{35}P(\s_5-\frac{3\pi i}{2})}&\non\\
&
+\frac{V_{3,1}(\s_5-\pi i)}{\s_{15}\s_{35}P(\s_3-\frac{\pi i}{2})}+
\frac{V_{3,2}(\s_5)}{(\s_{15}-\pi i)(\s_{35}-\pi i)P(\s_5+\frac{5\pi i}{2})}
&\non\\
&
+\frac{\tilde V_{2,0}(\s_5)}{\s_{15}\s_{35}P(\s_5+\frac{\pi i}{2})}+
\frac{\tilde V_{2,1}(\s_5)}{\s_{15}\s_{35}P(\s_5+\frac{3\pi i}{2})}
&
\label{f2n=6}
\end{align}
Let us note that the function $f_1(\s_1|\s_3|\s_4,\s_5)$ is
anti-symmetric w.r.t. the transposition 
of the last two variables $\s_4\leftrightarrow\s_5$
while the function $f_2(\s_1,\s_3|\s_5)$ is symmetric w.r.t.
the transposition of the first two variables $\s_1\leftrightarrow\s_3$.

Let us also note that the function $f_1$ defined by (\ref{f1n=6}) has
a similar to the case $n=5$ property, namely, that the only
non-zero residues for the pair of the variables $\s_3,\s_4$
are as follows
\begin{align}
&\mbox{Res}_{\s_4=\s_3+3\pi i} f_1(\s_1|\s_3|\s_4,\s_5)=
\chi_3(\s_1|\s_3|\s_5)&\non\\
&\mbox{Res}_{\s_4=\s_3+\pi i} f_1(\s_1|\s_3|\s_4,\s_5)=
\chi_1(\s_1|\s_3|\s_5)&\non\\
&\mbox{Res}_{\s_4=\s_3-\pi i} f_1(\s_1|\s_3|\s_4,\s_5)=
\chi_3(\s_1|\s_3-\pi i|\s_5)&
\label{f_1propn=6}
\end{align}
where the functions $\chi_1$ and $\chi_3$ may be calculated 
explicitly.
Of course, such a property is true for the pair of variables $\s_3,\s_5$ also.

Similar to the case $n=5$ arguments can be used in order to check
that the expressions (\ref{h16}) and (\ref{h26}) are well-defined
because due to the property (\ref{f_1propn=6}) a similar to (\ref{f_1prop2})
relation is valid also 
\begin{align}
&R_1\biggl(D_1f_1(\s_4|\s_1|\s_2,\s_3)-D_2f_1(\s_4|\s_2|\s_1,\s_3)\biggr)
=\d_{\s_1}\biggl(
\frac{\chi_1(\s_4|\s_1-\pi i|\s_3)}
{P(\s_1-\frac{\pi i}{2})P(\s_1+\frac{\pi i}{2})}
&\non\\
&
+
\frac{\chi_3(\s_4|\s_1-\pi i|\s_3)}
{P(\s_1-\frac{\pi i}{2})P(\s_1+\frac{3\pi i}{2})}
+
\frac{\chi_3(\s_4|\s_1-2\pi i|\s_3)}
{P(\s_1-\frac{3\pi i}{2})P(\s_1+\frac{\pi i}{2})}\biggr)&
\label{f_1prop2n=6}
\end{align}

In order to check the skew-symmetry of the answer (\ref{n=6}) one
can follow the same way like for the previous case $n=5$. All
necessary
relations may be straightforwardly generalized for the case $n=6$ also.
For instance, the integral formula (\ref{D2res}) may be generalized
as follows
\begin{align}
&
\int_{\Gamma_3}\frac{d\s_3}{2\pi i}
X_3(\s_3)\Phi(\s_3)D_3f_1(\s_4|\s_3|\s_1,\s_2)&\non\\
&
=D_1\biggl(p_3(\s_1)
\frac{\chi_3(\s_4|\s_1-\pi i|\s_2)}{P(\s_1-\frac{\pi i}{2})}-
p_3(\s_1+2\pi i)\frac{\chi_3(\s_4|\s_1|\s_2)}{P(\s_1+\frac{5\pi i}{2})}\biggr)
&\non\\
&
-D_2\biggl(p_3(\s_2)
\frac{\chi_3(\s_4|\s_2-\pi i|\s_1)}{P(\s_2-\frac{\pi i}{2})}-
p_3(\s_2+2\pi i)\frac{\chi_3(\s_4|\s_2|\s_1)}{P(\s_2+\frac{5\pi i}{2})}\biggr)
&\non\\
&
-p_3(\s_1)
\d^{-1}_{\s_1}\biggl[\frac{\mbox{Res}_{\s3=\s1+\pi i}}{P(\s_1+\frac{\pi i}{2})}
\biggl(D_1f_1(\s_4|\s_1|\s_3,\s_2)-
D_3f_1(\s_4|\s_3|\s_1,\s_2)\biggr)\biggr]
&\non\\
&
-p_3(\s_2)
\d^{-1}_{\s_2}\biggl[\frac{\mbox{Res}_{\s3=\s2+\pi i}}{P(\s_2+\frac{\pi i}{2})}
\biggl(D_2f_1(\s_4|\s_2|\s_1,\s_3)-
D_3f_1(\s_4|\s_3|\s_1,\s_2)\biggr)\biggr]
&
\label{D3resn=6}
\end{align}
Only one complication in comparison with $n=5$ appears because
of the second term with $D_2(\ldots)$ 
in the r.h.s. of (\ref{D3resn=6}) that  
makes a non-zero contribution after further integration. 
Because of this fact we had to include an additional terms
with the ratio $P(\s_3+\frac{3\pi i}{2})/P(\s_3+\frac{7\pi i}{2})$
in (\ref{q}) which appeared to be necessary for the correct definition
of the function $f_1$.

\end{document}